\documentclass[useAMS,usenatbib]{mn2e}
\usepackage{graphicx}
\usepackage{amssymb, amsmath}
\usepackage{array}

\def\fenh{\ensuremath{f_{\mathrm{enh}}}}

\newcommand{\ME}{\ensuremath{\mathrm{M_{\oplus}}}}

\newcolumntype{x}[1]{%
>{\raggedleft\hspace{0pt}}p{#1}}%

\title[Planet formation in optically thick discs]{Global models of planetary system formation 
in radiatively-inefficient protoplanetary discs}
\author[P.~Hellary and R.~P.~Nelson]
  {Phil Hellary$^{1}$\thanks{E-mail: p.hellary@qmul.ac.uk}
  and Richard~P.~Nelson$^{1}$\\ $^{1}$Astronomy Unit,
  Queen Mary University of London, Mile End
  Road, London, UK E1 4NS}
\begin{document}

\date{Accepted XXX.  Received XXX.}

\pagerange{\pageref{firstpage}--\pageref{lastpage}} \pubyear{2011}

\maketitle

\label{firstpage}

\begin{abstract}
{We present the results of N-body simulations of planetary systems formation
in radiatively-inefficient disc models, where positive corotation torques
may counter the rapid inward migration of low mass planets driven by
Lindblad torques. The aim of this work is to examine the nature of planetary
systems that arise from oligarchic growth in such discs. }
{We adapt the commonly-used Mercury-6 symplectic integrator by including
simple prescriptions for planetary migration (types I and II), planetary
atmospheres that enhance the probability of planetesimal accretion
by protoplanets, gas accretion onto forming planetary cores, and
gas disc dispersal. We perform a suite of simulations for a variety of
disc models with power-law surface density and temperature profiles,
with a focus on models in which unsaturated corotation torques 
can drive outward migration of protoplanets. In some models we account
for the quenching of corotation torques that arises when planetary
orbits become eccentric.}
{Approximately half of our simulations lead to the successful formation 
of gas giant planets with a broad range of masses and semimajor axes. 
Giant planet masses range from being approximately equal to that
of Saturn, up to approximately twice that of Jupiter. The semimajor
axes of these range from being $\sim 0.2$ AU, up to 
$\sim 75$ AU, with disc models that drive stronger outward migration 
favouring the formation of longer-period giant planets. Out of 
a total of 20 giant planets being formed in our simulation suite, we obtain 
3 systems that contain two giants. No super-Earth or Neptune-mass planets 
were present in the final stages of our simulations, in contrast to
the large abundance of such objects being discovered in observation surveys.
This result arises because of rapid inward migration suffered by massive
planetary cores that form early in the disc life time 
(for which the corotation torques saturate), combined with
gas accretion onto massive cores that leads them to become gas giants.
Numerous low mass planets are formed and survive in the simulations,
with masses ranging from a few tenths of an Earth mass, up to $\sim 3$ 
Earth masses. Simulations in which the quenching of corotation torques
for planets on modestly eccentric orbits was included failed to produce
any giant planets, apparently because Lindblad torques induce rapid
inward migration of planetary cores in these systems.} 
    
    
{We conclude that convergent migration induced by corotation torques 
operating during planet formation can enhance the growth rate of 
planetary cores, but these often migrate into the central star because 
corotation torques saturate. Outward migration of planetary
cores of modest mass can lead to the formation of gas giant planets
at large distances from the central star, similar to those observed
recently through direct imaging surveys. The excitation of planetary
eccentricities through planet-planet scattering during oligarchic growth 
may quench the effects of corotation torques, however, such that 
inward migration is driven by Lindblad torques.
}
\end{abstract}


\section{Introduction}
\label{section:intro}

Observations of extrasolar planets are providing a picture of a highly 
diverse population of bodies orbiting main sequence stars. 
At the extreme ends of the distribution, there exist very short-period
low-mass rocky planets such as CoRoT-7b \citep{Leger2009} and Kepler-10b 
\citep{Batalha2011}, and very long-period massive gas giant planets detected 
in recent years by direct imaging \citep{Marois, Kalas, Lagrange}.
In addition, there have been discoveries of short-period hot-Jupiters 
such as 51-Pegb \citep{MayorQueloz}, hot-Neptunes such as Gliese 436b 
\citep{Butler2004}, and super-Earths such as Gliese 581c \citep{Bonfils2005}.
Multiple planet systems are common, examples being the five planet system 
orbiting the star 55 Cancri consisting of gas giants and Neptune-mass bodies
\citep{Fischer2008}, and the recently reported Kepler-11 system,
consisting of six nearly coplanar low-mass planets \citep{Lissauer2011}.
A question that needs to be addressed is whether or not a particular model
of planet formation, such as the core-accretion model, can explain this
broad diversity, appealing to variety in the planet forming environment
to explain the range of observed systems. Or is it the case that quite 
different physical processes are operating on different length and/or time 
scales within protoplanetary discs to form planets with very different
characteristics. For example, core-accretion operating on long-times scales
at relatively close locations to the central star to form short-period 
systems, and disc gravitational fragmentation occurring at large radii on
short-time scales to form long-period giant planets.

To begin addressing this question in detail, it is necessary to construct 
global models of planetary formation that allow for the formation of
multiple planet systems with a diversity of masses and orbital elements
(semimajor axes, eccentricities etc). In this paper, we present the results from
global models, that are based on the oligarchic growth scenario for planet
formation, that have been constructed using a symplectic
N-body code \citep{cham99}, in conjunction with simplified prescriptions 
for the gas disc model, planetary migration, capture of planetesimals, 
gas-envelope accretion, and disc dispersal on Myr time scales. One of our main 
objectives in this work is to examine how our current understanding of migration 
of low-mass planets influences the formation of planetary systems, with
particular emphasis on the corotation torques in radiatively-inefficient
discs \citep{PBCK2010, PBK2011}.
As we anticipate that this is the first paper in a series that will
examine this issue, the prescriptions we have adopted in this initial
study for a number of physical processes, such as gas accretion, are 
necessarily very simplified. They serve the useful purpose, however,
of enabling N-body simulations to be performed of planetary system formation 
that lead to a diversity of outcomes, and these can be used to guide future 
model developments.

There is a substantial body of previous work that has examined the role
of migration in the formation of planets using N-body simulations.
\cite{PapaloizouLarwood2000} examined planetary growth through 
planet-planet collisions using N-body simulations combined with
prescriptions for migration and eccentricity/inclination damping.
\cite{McNeilDuncan2005} and \cite{daisaka} examined the effect of type I migration on
terrestrial planet formation, and
\cite{FoggNelson2007,FoggNelson2009} examined the the influence
of type I migration on the formation of terrestrial
planets in the presence of migrating Jovian-mass planets.
\cite{TerquemPapaloizou2007} examined the formation of
hot-super-Earths and hot-Neptunes using N-body simulations 
with type I migration.
\cite{McNeilNelson2009, McNeilNelson2010} carried out large-scale
simulations of oligarchic growth to explore the formation of systems
of  hot-Neptunes and super-Earths, such those around the stars
Gliese 581 and HD698433, using a novel symplectic integrator with multiple 
timesteps. An alternative to these approaches has been planetary 
population synthesis modelling, as presented by
\cite{IdaLin2008, IdaLin2010}, \cite{Mordasini2009a},
\cite{Mordasini2009b}, and \cite{Miguel2011}.
These monte-carlo approaches have significant advantages in being
able to cover a very broad range of parameter space, allowing
meaningful statistical comparisons with observational data to
be undertaken. The computational efficiency also allows 
complex models of planetary atmospheres and gas accretion to be
incorporated, as presented by the Mordasini et al work. Accurate
treatment of planet-planet gravitational interactions are difficult
to include in these models, however, such that predictions about planetary system
multiplicity, orbital eccentricities and inclinations are not a natural
outcome of the models (we note that models by \cite{IdaLin2010} now include
a simplified treatment of planet-planet interaction dynamics).
 
This paper is organised as follows. In Sect.~\ref{method} we present the
numerical methods and the physical model. In Sect.~\ref{initial} we
present the initial conditions for the simulations. Results are
described and analysed in Sect.~\ref{endresults}, comparisons are made to observations in Sect.~\ref{realobs} and a discussion and
concluding remarks are provided in Sect.~\ref{conclusions}.

\section{Method}
\label{method}

In the following subsections we give details about our physical model
and the numerical methods we employ.
  
\subsection{Gas disc model}
\label{gasdisc}
To limit the parameter space covered by the simulations, we consider only disc 
models that can provide outward migration when unsaturated corotation torques are 
included. The conditions under which outward migration occurs are discussed in
later sections, but as a rule of thumb we find that the temperature radial 
power-law index, $\beta$, must be approximately 0.25 larger than the surface 
density power-law index, $\alpha$.
  
The gas surface density is given by the power-law expression
  \begin{equation}
   \Sigma_{g}(R,t)=\Sigma_{g}\left(1 \, {\rm AU}\right)
   \left(\frac{R}{1 \, {\rm AU}}\right)^{-\alpha} 
   \exp\left(-t/\tau_{\rm disc}\right)
    \label{eq:sigma}
    \end{equation} 
where the factor $\exp\left(-t/\tau_{\rm disc}\right)$ mimics
the dispersal of the gas disc by viscous evolution and photoevaporation
on a time scale defined by $\tau_{\rm disc}$.
The volume density of gas is then
\begin{equation}
 \rho(R,z,t) = \frac{\Sigma(R,t)}{\sqrt{2 \pi} H} \exp{(-z^2/2H^2)}
\label{eq:rho}
\end{equation}
where $H$ is the local disc scale height.
The disc temperature is also given by a power-law function of radius
  \begin{equation}
   T(R)=T\left( 1 \, {\rm AU} \right)
   \left(\frac{R}{1 \, {\rm AU}}\right)^{-\beta}.
  \end{equation} 
A disc with power-law density and temperature profiles also
has a power-law entropy profile. The associated power-law
index is given by
\begin{equation}
\zeta = (\alpha + \beta) - \alpha \gamma
\label{eq:zeta}
\end{equation}
where $\gamma$ is the usual ratio of specific heats,
here taken to be $\gamma=7/5$.
The isothermal sound speed is
  \begin{equation}
   c_s=\sqrt{\frac{k_{\rm B} T}{m_{\rm H} \mu}}
  \end{equation}
where $k_{\rm B}$ is the Boltzmann constant, $m_{\rm H}$ is 
the mass of a hydrogen atom and $\mu$ is the mean molecular weight
(here assumed to equal 2.4).
The disc scale height is given by
 \begin{equation}
   H=c_s\Omega_{\rm K}
 \end{equation}
where $\Omega_{\rm K}$ is the Keplerian angular velocity.
The angular velocity of the gas is given by
\begin{equation}
\Omega(R) = \Omega_K(R) \left[1 - (\alpha+\beta) 
            \left(\frac{H}{R}\right)^2 \right].
\label{eq:Omega}
\end{equation}

\subsection{Opacities}
\label{sec:opacity}

  We take the opacity $\kappa$ to be always equal to the Rosseland
  mean opacity, and we take the temperature and density dependence to
  be given by the formulae of \cite{Bell97} below 3730 K and by
  \cite{Bell94} above this value:
  \begin{equation}
    \label{eq:bellopacity}
    \kappa [\mathrm{cm^2/g}] = \left\{
      \begin{array}{ll}
        10^{-4} \,\, T^{2.1}   & T < 132 {\, \mathrm K} \\[.5em]
        3 \, T^{-0.01}   & 132 {\, \mathrm K} \leq T < 170 {\, \mathrm K} \\[.5em]
        T^{-1.1}   & 170 {\, \mathrm K} \leq T < 375 {\, \mathrm K} \\[.5em]
        5 \times 10^{4} \, T^{-1.5}   & 375 {\, \mathrm K} \leq T < 390 {\, \mathrm K} \\[.5em]
        0.1 \, T^{0.7}   & 390 {\, \mathrm K} \leq T < 580 {\, \mathrm K} \\[.5em]
        2 \times 10^{15}\, T^{-5.2}   & 580 {\, \mathrm K} \leq T < 680 {\, \mathrm K} \\[.5em]
        0.02 \, T^{0.8}   & 680 {\, \mathrm K} \leq T < 960 {\, \mathrm K} \\[.5em]
        2 \times 10^{81}\, \rho \, T^{-24}   & 960 {\, \mathrm K} \leq T < 1570 {\, \mathrm K} \\[.5em]
        10^{-8} \, \rho^{2/3} \, T^{3}   & 1570 {\, \mathrm K} \leq T < 3730 {\, \mathrm K} \\[.5em]
        10^{-36} \, \rho^{1/3} \, T^{10}   & 3730 {\, \mathrm K} \leq T < 10000 {\, \mathrm K} \\[.5em]

      \end{array}
    \right.
  \end{equation}

\subsection{Disc solid component}
\label{solid}
The disc solid component is composed initially of protoplanets
and planetesimals (that we model as a computationally feasible
number of `superplanetesimals' of much larger mass than real
planetesimals, but with an assumed radius equal to that of
realistic planetesimals (10 km) such that they experience the appropriate
gas drag force.)
Protoplanets are initially spaced by 10 mutual Hill radii, and
planetesimals are scattered throughout the disc such that the 
total solids content follows the surface density power-law 
prescribed for the gaseous component. 
As in \cite{thommes03}, planetesimals are distributed according to a 
Rayleigh distribution and have RMS values of the eccentricity $e=0.01$ and 
inclination $i=0.005$ radians, respectively. The surface density of solids 
is enhanced beyond the snow line, whose position $R_{\rm snow}$
is determined by the location where the temperature falls below 170 K. 
The snow line discontinuity is spread over a distance $\sim 1$ AU:
 \begin{equation}
  \Sigma_{s,0}(R)=\left\{\Sigma_1+\left(\Sigma_2-\Sigma_1\right)\left[\frac{1}{2}\left(\frac{R-R_{\rm snow}}{0.5 \, {\rm AU}}\right)+\frac{1}{2}\right]\right\}\left(\frac{R}{1 \, {\rm AU}}\right)^{-\alpha}
  \end{equation}
  
The surface density enhancement due to the snowline $(\Sigma_2/\Sigma_1)=
30/7.1$ as in \cite{thommes03}. Planetesimal densities are set at 
3 ${\mathrm{g/cm^3}}$ throughout the disk. Protoplanet densities are set 
at 3 ${\mathrm{g/cm^3}}$ inside the snowline and 1.5 
${\mathrm{g/cm^3}}$ beyond, as defined by \cite{thommes03}.
The mass of the protoplanets at $t=0$ is $m_{\rm p}=0.02$ M$_{\oplus}$,
and the mass of the superplanetesimals is 0.004 M$_{\oplus}$.

 \subsection{Aerodynamic drag}
  \label{aero}
 For kilometre-sized planetesimals, aerodynamic drag provides an
efficient source of eccentricity and inclination damping. 
We apply gas drag to all bodies in the simulation in the form of 
Stokes' drag law (\citealt{adachi}),
  \begin{equation}
  \pmb{F}_{\rm drag}=m_{\rm p}
\left(\frac{-3\rho C_{\rm D}}{8\rho_{\rm p} r_{\rm p}}\right)
v_{\rm rel} \pmb{v}_{\rm rel}
  \end{equation}
where $\rho$ is the local gas density, $\rho_{p}$ is the density of the planetesimal, $r_{\rm p}$ is the
radius of the body and $C_{\rm D}$ is a dimensionless drag
coefficient (here taken to be unity).

\subsection{Capture radii enhancement due to atmospheric drag}
\label{atmos}
If a protoplanet becomes large enough to accumulate a gaseous atmosphere, 
the gas drag acting on planetesimals passing through this atmosphere has 
the effect of increasing the effective capture radius. 
We use the prescription described in section 2.5 of \cite{inaba03} to 
model this effect (see their Eq. 17 and appendix A). In this model, a planetesimal that 
is within the Hill sphere of the protoplanet, and located a distance $r_{\rm c}$ 
from the centre of the protoplanet, will be captured if its physical radius is less than 
$r_{\rm crit}$ given by the expression
 \begin{equation}
   r_{\rm crit}=\frac{3}{2}\frac{v_{\rm rel}^2+2 G m_{\rm p}/r_{\rm c}}{v_{\rm rel}^2+2G m_{\rm p}/r_{\rm H}}\frac{\rho(r_{\rm c})}{\rho_{\rm p}}.
    \end{equation}
Here $\rho(r_{\rm c})$ is the local density of the protoplanet 
atmosphere, $\rho_{\rm p}$ is the density of the planetesimal, 
$r_{\rm H}$ is the protoplanet's Hill radius and $v_{\rm rel}$ 
is the relative velocity between the two bodies.

The atmosphere model requires us to specify the luminosity of the
planet. We assume that this is equal to the gravitational energy released 
by incoming planetesimals
 \begin{equation}
  L_p=
\frac{Gm_{\rm p}}{r_{\rm p}}\frac{dm_{\rm p}}{dt}
\end{equation}

We monitor the accretion rate of solids experienced by protoplanets in our 
simulations, and use this to determine the accretion luminosity.
In order to smooth out the stochastic nature of planetesimal accretion, 
we calculate and use the average luminosity of a protoplanet over 
temporal windows of 200 local orbits (or 4000 years, whichever is smaller). 
We limit the calculated luminosity to within the range $10^{-9}$ to $10^{-4}$ 
L$_{\odot}$.

\begin{figure}
      \includegraphics[width=80mm, clip]{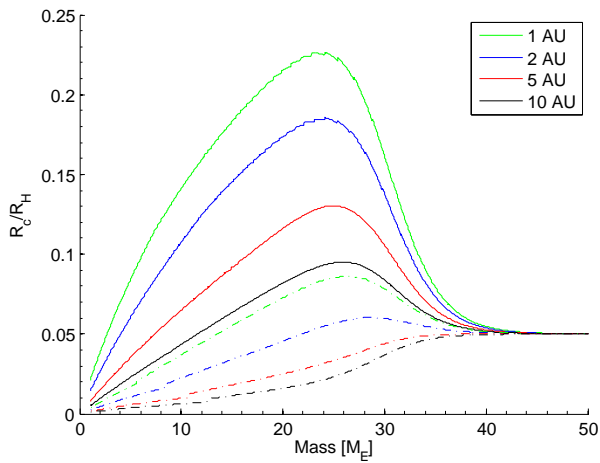}
      \caption{Effective planetesimal capture radius enhancement due to atmospheric 
drag for 10 km planetesimals and with various planet luminosities.
Solid lines correspond to $L_{\rm p}=10^{-8}$ L$_{\odot}$; dot-dashed lines
correspond to $L_{\rm p}=10^{-5}$ L$_{\odot}$.\label{plot:atmosphericdrag} }
  \end{figure}

The \cite{inaba03} model assumes that the contribution to the gravitating
mass from the atmosphere is negligible compared to that of the solid core. 
In order to avoid an overestimation of the capture radius of larger 
planets, we limit the effective capture radius of a 
planet to be a maximum of $1/20$ of a planet's Hill radius for 
these planets. The transition is smoothed using the expression
\begin{equation}
    \begin{split}
    r_{\rm capture}=
    \left[0.5-0.5\tanh{\left(\frac{m_{\rm p}-30 M_{\oplus}}{5 M_{\oplus}}
    \right)}\right] r_{\rm atmos} \\
+\left[0.5+0.5 \tanh{\left(\frac{m_{\rm p}-30M_{\oplus}}{5M_{\oplus}}
\right)}\right]0.05 r_{\rm H}
\end{split}
\end{equation}
where $r_{\rm capture}$ is the effective capture radius, $r_{\rm atmos}$ 
is the enhanced capture radius due to the atmosphere and 
$r_{\rm H}$ is the Hill radius. 
Figure \ref{plot:atmosphericdrag} shows the effective capture radius as
a function of planet mass and luminosity, including the above smoothing
procedure.
  
\subsection{Type I migration in radiatively-inefficient discs}
\label{noniso}
We include the migration of low mass planets in our simulations
by implementing the torque formulae presented by \cite{PBCK2010} and \cite{PBK2011}. 
These formulae describe how the various torque contributions vary as 
the planet mass and local conditions in the disc change. Specifically, 
corotation torques depend sensitively on the ratio of the horseshoe 
libration time scale to either the viscous or thermal diffusion 
time scales.

There are two basic contributions to the corotation torque: the
vorticity-related corotation torque and the entropy-related
corotation torque. In an inviscid two dimensional disc, the vortensity
(ratio of vorticity to surface density) is a conserved quantity 
on streamlines.
Fluid elements undergoing horseshoe orbits in the presence of a planet
therefore conserve this quantity. For power-law surface density profiles 
with indices greater (less negative) than $-3/2$, there is a negative radial 
vortensity gradient, and the exchange of angular momentum between an 
embedded planet and disc material as the fluid follows horseshoe streamlines 
generating a positive torque on the planet \citep{GoldreichTremaine1979}. 
In the absence of viscous diffusion, 
material undergoing horseshoe orbits eventually phase mixes because of the varying 
horseshoe orbit time scales, erasing the vortensity 
profile in the corotation region and saturating the corotation torque
(i.e. switching it off).
The action of viscous stresses can desaturate the horseshoe torque by
maintaining the vortensity gradient across the horseshoe region, and this
occurs optimally when the viscous diffusion time scale across the width of
the horseshoe region is approximately equal to half the horseshoe libration time.
The presence of horseshoe streamlines inevitably means that the associated
horseshoe torque is a non linear effect (because horseshoe orbits are
not present in a linear theory), usually referred to as horseshoe
drag \citep{Ward1991, Masset2001, PaardekooperPapaloizou2008}. As the viscosity is 
increased above its optimal value the vortensity on streamlines begins to 
be modified significantly as the
fluid undergoes horseshoe U-turns. For large enough viscosity the 
vorticity-related corotation torque 
eventually approaches the smaller value obtained from linear theory
\citep{Masset2002}.
This arises when the viscous diffusion time is shorter than the horseshoe
U-turn time.

A similar process occurs for the entropy-related corotation torque,
but in this case the controlling parameter is the thermal diffusion
time scale instead of the viscous one. Optimal corotation torques 
are obtained when both the thermal and viscous diffusion time scales 
across the width of the horseshoe region are equal to approximately half 
the horseshoe libration time. It should be noted that, in addition
to thermal diffusion, viscosity is required to desaturate the entropy-related
horseshoe torque. This is because material trapped in the horseshoe region
in an inviscid disc constitutes a closed system that can only exchange
a finite quantity of angular momentum with the planet. Viscosity is
required to couple this region with the rest of the disc, such that 
exchange of angular momentum can desaturate the corotation torque.

For simplicity of implementation we adopt the approximation suggested by 
\cite{LyraPaardekooper2010} and assume that the thermal and viscous 
time scales in the disc are equal. For a disc in thermodynamic equilibrium, 
where the heating is provided by viscous dissipation and local cooling 
is via blackbody radiation, this is a reasonable assumption to make. 

Based on the above discussion, the torque experienced by a low mass planet
embedded in a disc depends on the Lindblad torques (originating from
the excitation of density waves at Lindblad resonances), and a weighted
sum of the vorticity-related horseshoe drag, the entropy-related horseshoe
drag, the vorticity-related linear corotation torque, and the 
entropy-related linear corotation torque. 
These torque contributions are given as follows: \\
The Lindblad torque is
\begin{equation}
\Gamma_{\rm LR} = \left(\frac{\Gamma_0}{\gamma}\right) 
\left[-2.5 -1.7 \beta +0.1 \alpha \right]
\label{lindblad-torque}
\end{equation}
the vorticity-related horseshoe drag is
\begin{equation}
\Gamma_{\rm VHS} = \left(\frac{\Gamma_0}{\gamma}\right) 
\left[ 1.1 \left(\frac{3}{2} - \alpha \right) \right] 
\label{vort-horseshoe-drag}
\end{equation}
the entropy-related horseshoe drag is
\begin{equation}
\Gamma_{\rm EHS} = \left(\frac{\Gamma_0}{\gamma}\right) 
\left[ 7.9 \frac{\zeta}{\gamma} \right]
\label{ent-horseshoe-drag}
\end{equation}
the vorticity-related linear corotation torque is
\begin{equation}
\Gamma_{\rm LVCT} = \left(\frac{\Gamma_0}{\gamma}\right) 
\left[ 0.7\left(\frac{3}{2} - \alpha \right) \right]
\label{vort-linear-CT}
\end{equation}
the entropy-related linear corotation torque is
\begin{equation}
\Gamma_{\rm LECT} =  \left(\frac{\Gamma_0}{\gamma}\right) 
\left[ \left(2.2 -\frac{1.4}{\gamma}\right) \zeta \right]
\label{ent-linear-CT}
\end{equation}
In these above expression $\gamma$ is the ratio of specific heats,
and $\Gamma_0$ is given by 
  \begin{equation}
  \Gamma_0=
\left(\frac{m_{\rm p}}{M_*}\right) 
\left(\frac{m_{\rm p}}{\Sigma_{\rm p} a_{\rm p}^2} \right)
\left(\frac{a_{\rm p} \Omega_{\rm p}}{c_s}\right)^2
a_{\rm p}^2\Omega_{\rm p}^2,
\label{eq:Gamma0}
  \end{equation}
where $a_{\rm p}$ is the planet semimajor axis, and a subscript `p'
denotes that quantities should be evaluated at the orbital location
of the planet.
In order to obtain the correct total torque as a function of the
thermal and viscous diffusion coefficients we combined the
above individual torque expressions into the following
formula \citep{PBK2011}:
\begin{eqnarray}
\Gamma_{\rm tot} & = & \Gamma_{\rm LR} + \left\{\Gamma_{\rm VHS} F_{\nu} G_{\nu} 
+\Gamma_{\rm EHS} F_{\nu} F_{\rm d} \sqrt{G_{\nu} G_{\rm d}}  \right. \\ \nonumber
& + & \Bigl. \Gamma_{\rm LVCT} (1-K_{\nu}) +\Gamma_{\rm LECT} (1-K_{\nu}) (1-K_{\rm d})
\Bigr\} E
\label{total-torque}
\end{eqnarray}
where the functions $G_{\rm v}$, $G_{\rm d}$, $F_{\rm v}$, $F_{\rm d}$,
$K_{\rm v}$ and $K_{\rm d}$ are related either to the ratio between the
viscous/thermal diffusion time scales and the horseshoe libration
time scale, or to the ratio of the viscous/thermal diffusion time scales 
and the horseshoe U-turn time scale.  The factor $E$, that multiplies 
all terms that can contribute to the corotation torque, allows for
the fact that corotation torques may be strongy attenuated when
the planet has a finite eccentricity, such that it undergoes radial
excursions that are larger than the width of the horseshoe region
\citep{BitschKley2010}.
To account for this effect we define $E$ according to
\begin{equation}
E = (1 - \tanh{(e/x_{\rm s})}.
\label{eq:E}
\end{equation}
where the dimensionless horseshoe width is given by
\begin{equation}
x_{\rm s} = \frac{1.1}{\gamma^{1/4}} \sqrt{\frac{q}{h}},
\label{eq:x_s}
\end{equation}
$q=m_{\rm p}/M_*$ and $h=H/R$. Note that for most simulations we set $E=1$, but for a 
subsample of our runs (labelled as `E') we use Eq.~(\ref{eq:E}) to 
define $E$.

The horseshoe libration time is given by 
$t_{\rm lib}=8 \pi/(3 \Omega_{\rm p} x_{\rm s})$, and
the viscous diffusion time scale across the horseshoe region is
given by $t_{\rm v}=(x_{\rm s} a_{\rm p})^2/\nu$, where $\nu$ is the viscous
diffusion coefficient. Similarly the thermal diffusion time scale is
given by $t_{\rm d}=(x_{\rm s} a_{\rm p})^2/D$, where $D$ is the
thermal diffusion coefficient (defined below).
Following \cite{PBK2011}, we define two parameters that determine the relation between
the thermal/diffusion time scales and the horseshoe libration time scale
\begin{equation}
p_{\rm v} = \frac{2}{3} \sqrt{\frac{a_{\rm p}^2 \Omega x_{\rm s}^3 }{ 2 \pi \nu}}
\equiv \sqrt{\frac{16}{27} \frac{t_{\rm v}}{t_{\rm lib}}},
\label{pv}
\end{equation}
which we refer to as the viscous diffusion parameter
\begin{equation}
p_{\rm d} = \sqrt{\frac{a_{\rm p}^2 \Omega x_{\rm s}^3 }{ 2 \pi D}} 
\equiv \sqrt{\frac{4}{3} \frac{t_{\rm d}}{t_{\rm lib}}},
\label{pd}
\end{equation}
which we refer to as the thermal diffusion parameter
Note that $\nu$ and $D$ are assumed to be equal in this work,
and are complicated functions of radial position in the
disc because of the functional form used to define the
opacity in Eq.~(\ref{eq:bellopacity}).
These diffusion parameters are used to define the following functions
\begin{equation}
F_{\rm v} = \frac{1}{\left[1+(p_{\rm v}/1.3)^2\right]}
\label{Fv}
\end{equation}
\begin{equation}
F_{\rm d} = \frac{1}{\left[1+(p_{\rm d}/1.3)^2\right].}
\label{Fd}
\end{equation}
Using the viscous diffusion parameter $p_{\rm v}$ we also define the
following functions
\begin{equation}
G_{\rm v} =
\begin{cases}
\frac{16}{25} \left( \frac{45 \pi}{8}\right)^{3/4} p_{\rm v}^{3/2}
&  \text{if $p_{\rm v} < \sqrt{8/(45 \pi)}$} \\ \\
1 - \frac{9}{25} \left(\frac{8}{45 \pi}\right)^{4/3} p_{\rm v}^{-8/3}
& \text{if $p_{\rm v} \ge \sqrt{8/(45 \pi)}$}
\end{cases}
\label{eq:G_v}
\end{equation}
\begin{equation}
K_{\rm v} =
\begin{cases}
\frac{16}{25} \left( \frac{45 \pi}{28}\right)^{3/4} p_{\rm v}^{3/2}
& \text{if $p_{\rm v} < \sqrt{8/(45 \pi)}$} \\ \\
1 - \frac{9}{25} \left(\frac{28}{45 \pi}\right)^{4/3} p_{\rm v}^{-8/3}
& \text{if $p_{\rm v} \ge \sqrt{28/(45 \pi)}$}
\end{cases}
\label{eq:K_v}
\end{equation}
Using the thermal diffusion parameter $p_{\rm d}$ we define
the following functions
\begin{equation}
G_{\rm d}= 
\begin{cases}
\frac{16}{25} \left( \frac{45 \pi}{8}\right)^{3/4} p_{\rm d}^{3/2}
& \text{if $p_{\rm d} < \sqrt{8/(45 \pi)}$} \\ \\
1 - \frac{9}{25} \left(\frac{8}{45 \pi}\right)^{4/3} p_{\rm d}^{-8/3}
& \text{if $ p_{\rm d} \ge \sqrt{8/(45 \pi)}$}
\end{cases}
\label{eq:G_d}
\end{equation}
\begin{equation}
K_{\rm d} = 
\begin{cases}
\frac{16}{25} \left( \frac{45 \pi}{28}\right)^{3/4} p_{\rm d}^{3/2}
& \text{if $p_{\rm d} < \sqrt{28/(45 \pi)}$} \\ \\
1 - \frac{9}{25} \left(\frac{28}{45 \pi}\right)^{4/3} p_{\rm d}^{-8/3}
& \text{if $p_{\rm d} \ge \sqrt{28/(45 \pi)}$}
\end{cases}
\label{eq:K_d}
\end{equation}

\subsubsection{Thermal and viscous diffusion}
Radiative diffusion in the disc causes the thermal energy per unit volume, $e$,
to evolve according to
\begin{equation}
\frac{\partial  e}{\partial t} = - \nabla . {\bf F}
\label{eq:heat-diff}
\end{equation}
where the radiative flux in the radial direction (across the horseshoe region)
may be expressed as
\begin{equation}
F_r = - \frac{4 a_{\rm r} c }{3} \frac{T^3}{\kappa \rho} \frac{dT}{dr}.
\label{eq:flux}
\end{equation}
Here $a_{\rm r}$ is the radiation constant and $c$ is the speed of light.
Noting that $e=P/(\gamma -1)$ and $P=k_{\rm B} \rho T/ (\mu m_{\rm H}$), 
and assuming that $\rho$ is locally constant, we obtain the diffusion equation 
governing temperature evolution
\begin{equation}
\frac{\partial T}{\partial t} = \nabla_r \left(D \frac{dT}{dr}\right)
\label{eq-T-diff}
\end{equation}
where the diffusion coefficient, $D$, is given by
\begin{equation}
D = \frac{4 a_{\rm r} c}{3} \frac{T^3}{\kappa \rho^2} 
\frac{(\gamma-1) \mu m_H}{k_{\rm B}}.
\label{eq:D-coeff}
\end{equation}
We set the viscous diffusion coefficient equal to the
thermal diffusion coefficient for the purpose
of determining the level of saturation of
corotation torques ($\nu = D$).

\subsubsection{Eccentricity and inclination damping}
To damp the inclinations of protoplanets we used the prescription 
given in Appendix A of \cite{daisaka}:
  \begin{equation}
  F_{idamp,z}=m_p
\left(\frac{m_p}{M_{\odot}}\right)
\left(\frac{a_p\Omega_p}{c_s}\right)^4
\left(\frac{\Sigma_ga_p^2}{M_{\odot}}\right)
\Omega_p\left(2A_z^cv_z+A_z^sz\Omega_p\right)
\end{equation}
where $A_z^c=-1.088$ and $A_z^s=-0.871$.

To damp eccentricities we used a simple time scale damping formula given by
\begin{equation}
 F_{edamp,r}=-\frac{v_r}{t_{edamp}} , \,\,\, 
 F_{edamp,\theta}=-\frac{0.5\left(v_\theta-v_K\right)}{t_{edamp}}
 \end{equation}
 where
  \begin{equation}
  t_{edamp}=
\left(\frac{m_p}{M_{\odot}}\right)^{-1}
\left(\frac{a_p\Omega_p}{c_s}\right)^{-4}
\left(\frac{\Sigma_ga_p^2}{M_{\odot}}\right)^{-1}
\Omega_p^{-1}
  \end{equation}
This prescription was adopted rather than using the eccentricity damping
forces prescribed in \cite{daisaka} because we found that they
could generate significant jitter in the acceleration experienced
by the planets in disc models with strong radial temperature gradients,
where $H/r$ becomes small near the disc outer edge. The formula based
on the time scale argument produced smoother results, apparently
because it is based on an orbit-averaging procedure rather than capturing 
the instantaneous force experienced by a planet around its orbit.
  
  \subsection{Gas envelope accretion}
  \label{gasenvelope}
 As protoplanets grow through mutual collision and planetesimal accretion
 they are able to accrete a gaseous envelope from the 
 surrounding disc, and may eventually become gas giant planets.
 To model gaseous envelope accretion, we have implemented a 
 very approximate scheme by calculating fits to the results of
 1D giant planet formation calculations presented by 
 \cite{movshovitz}. 
 Working in time units of Myr and mass units of Earth masses, 
the gas accretion rate is given by
  \begin{equation}
  \frac{dm_{\rm ge}}{dt}=5.5\left(\frac{1}{\tau_{\rm ge}-\tau_{\rm ge0}}
  \right)
  \end{equation}
  where we define $\tau_{\rm ge}$ by the expression 
  \begin{equation}
  \tau_{\rm ge}=9.665M_{\rm core}^{-1.2},
  \end{equation}
  and $\tau_{\rm ge0}$ is given by the expression
  \begin{equation}
  \tau_{\rm ge0}=\tau_{\rm ge}\left(1-\left(\frac{1}{\exp{\left(
  \frac{m_{\rm ge}}{5.5}\right)}}\right)\right).
  \end{equation}
   \begin{figure}
      \includegraphics[width=80mm, clip]{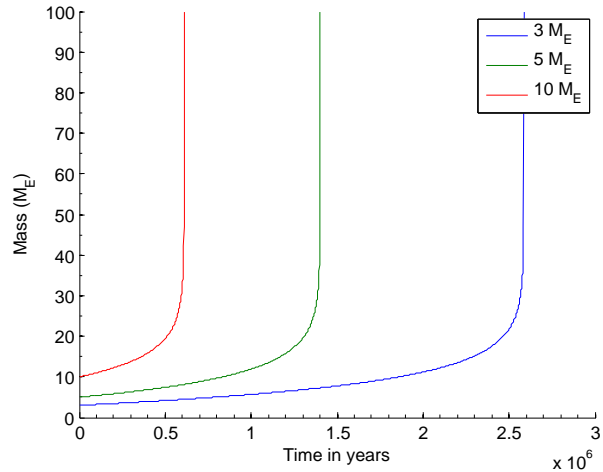}
      \caption{Gas accretion onto giant planet cores for 3, 5 and 10 $M_{\oplus}$ cores against time 
               at 5 AU in a disc with no migration or planetesimal accretion.
               \label{plot:massgrowth} }
  \end{figure}
This procedure allows the planet core to grow due to accretion of solids after envelope accretion 
has commenced, and allows the rate of envelope accretion to adjust to the changing core mass.
It is well known from the studies of \cite{pollack96} and others that the rate of gas envelope
accretion is a sensitive function of the core mass, increasing as the core mass increases.
Figure \ref{plot:massgrowth} shows the gas envelope mass evolution in the absence of planetesimal 
accretion and migration for planets with fixed core mass. These are very similar to the results 
of detailed 1D giant planet formation calculations displayed in figure 1 of \cite{movshovitz}. 
Although we have implemented the above equations for gas accretion numerically, we
note that they have the analytic solution
\begin{equation}
m_{\rm ge}(t) = - 5.5 \ln{\left(1 - \frac{m_{\rm core}^{1.2}}{9.665} t\right) }.
\label{eq:m_ge_analytic}
\end{equation}
Ideally, we would like to incorporate full 1D models of giant planet formation
in our N-body simulations, such that the gas envelope accretion can respond to the
changing planetesimal accretion rate and changing conditions in the disc, but such an approach 
is computationally prohibitive at present. Our simplified approach is highly efficient and provides
a reasonably good approximation to detailed core nucleated accretion models, enabling us to 
add a vital missing component to our N-body simulations.

The amount of gas that can accrete rapidly onto a forming giant planet is constrained
by the availability of gas in the local feeding zone. We allow giant planets
to accrete gas using the procedure described above until the envelope mass approaches
the isolation mass, defined to be the gas mass in the feeding zone.
We approximate the feeding zone width to be \citep{Lissauer1993}
  \begin{equation}
  r_{\rm c}=2\sqrt3 r_{\rm H}
  \end{equation}
 leading to the following expression for the gas isolation mass of the planet 
  \begin{equation}
  m_{\rm g-iso}=\int_{a-2\sqrt{3}r_{\rm H}}^{a+2\sqrt{3}r_{\rm H}}2\pi\Sigma_{\rm g} R \, dR.
  \label{eq:g-iso}
  \end{equation}

During the growth of the planet, it can begin to open a gap through nonlinear
tidal interaction with the disc \citep{LinPapaloizou1986} 
and for typical disc parameters this occurs around a Jovian mass
\citep{Bryden1999, Kley1999, Nelson2000}. Consequently,
if the isolation mass exceeds the Jovian mass, we limit the mass
of the planet that can be obtained during runaway gas accretion to be the 
Jovian mass.

Once the runaway cut-off mass has been reached, the gas accretion rate switches to the rate 
that viscosity can supply mass to the planet from the gas disc,
  \begin{equation}
  \frac{dm_{\rm ge}}{dt}=3\pi\nu\Sigma_{\rm g}
  \end{equation}
where $\nu$ is the local disc viscosity given by
  \begin{equation}
  \nu=\alpha_{\rm v} H^2\Omega(R)
  \label{eq:nu}
  \end{equation}
where $\alpha_{\rm v}$ is the viscous parameter (set to $10^{-3}$ for the purpose
of this calculation). Note that this value for the kinematic viscosity is not
the same as that obtained by assuming the thermal and viscous diffusion coefficients
are equal, as is done to determine the magnitude of the corotation torques acting on
a planet (see Sect.~\ref{noniso}). However, the value of $\alpha_{\rm v}$ adopted for the 
purpose of determining the viscously-driven mass accretion rate is similar to that 
used in many previous studies of disc-planet interactions 
\citep{Bryden1999, Kley1999, Nelson2000}, and produces viscous accretion rates within
the range observed to occur onto T Tauri stars \citep{Sicilia-Aguilar}.

  \subsection{Type II migration}
  \label{typeii}
For massive planets, the migration changes from being of type I 
to being of type II as gap formation occurs. Under these circumstances the planet
migrates inward on a time scale equal to the local viscous evolution time, $\tau_{\nu}$,
provided that the planet mass is smaller than the local disc mass. For more massive
planets the migration slows down due to the inertia of the planet 
(and is ultimately determined by the time over which the viscous flow in the disc delivers 
a mass of gas comparable to that of the planet to the planet orbital radius 
\citep{Ivanov, SyerClarke}).

The viscous evolution time is $\tau_{\nu} = R^2/(3 \nu)$, where we use Eqn.~\ref{eq:nu}
to calculate $\nu$, and we apply the following torque in the type II migration regime
\begin{equation}
\Gamma_{\rm II} = -\frac{m_{\rm p} j_{\rm p}}{\tau_{\nu}} 
\left(1+\frac{m_{\rm p}}{\pi a_{\rm p}^2 \Sigma_{\rm p}}\right)^{-1}
\label{eq:typeII}
\end{equation}
where $m_{\rm p}$ is the planet mass, $j_{\rm p}$ is the specific angular momentum,
$a_{\rm p}$ is the planet semimajor axis and $\Sigma_{\rm p}$ is the disc surface
density at the planet's semimajor axis location.
We transition smoothly between the type I and type II migration regimes
using the following expression
\begin{equation}
\Gamma_{\rm eff} = \Gamma_{\rm II} B + \Gamma_{\rm I} (1 - B)
\label{eq:smooth}
\end{equation}
where $\Gamma_{\rm eff}$ is the torque applied during the transition,
$\Gamma_{\rm I}$ is the type I torque, and the transition function $B$ is
given by
\begin{equation}
 B=0.5+0.5\tanh\left(\frac{m_p-65 M_{\oplus}}{15 M_{\oplus}}\right).
\end{equation}
This form for $B$ was adopted to allow the transition to 
type II migration to begin for $m_{\rm p}=30 \ME$, and for the
transition to be complete for planets with mass $m_{\rm p}=100 \ME$,
in broad agreement with the results from analytic considerations
\citep{ward97} and numerical simulations \citep{dangelo-kley}.
In the type II regime, eccentricities and inclinations are damped
on a time scale equal to $\tau_{\nu}/10$.

\section{Initial conditions}
\label{initial}

Our simulations were performed using the Mercury-6 symplectic integrator \citep{cham99}, 
modified to include the physics described in Sect. 2. 
In order to model feasibly multiple parameter sets over time scales of 3 Myr, our 
planetesimal disc consists of super-planetesimals (0.004 M$_{\oplus}$) with effective 
radius of 10km, representing the averaged orbits of a much larger number of real 
planetesimals. We set the mass of our protoplanets to be a factor of 5 larger 
than the planetesimals, giving run times of approximately three cpu weeks for each 
simulation.

To enable broad coverage of the $\alpha$ and $\beta$ parameter set,
we limited the number of realisations of initial conditions to two runs for each parameter choice,
with each member of the pair differing only by the random number 
seed used to determine initial positions of the planetesimals.
Our initial suite of simulations included models with enhancements by factors of 3 and 5 
above the mass of the Minimum Mass Solar Nebula (MMSN) \citep{hayashi} (models labelled `M'), but we later augmented these with additional
models with mass enhancement factors equal to 1 and 2 (models labelled `R'). We also examined
two models where we implemented a reduction  in the corotation torques for
protoplanets that develop eccentric orbits (discussed in detail in Sect.\ref{noniso}). 
These models are labelled `E'. Test calculations examining the influence of the
planetesimal capture radii of protoplanets were also performed.

In order to ensure that the disc mass is locally comparable in models with different
surface density profiles, we normalise the disc masses so that they all
have the same mass in the region from 2-15 AU that the enhanced MMSN discs would have.
This resulted in there being 28 protoplanets, with $\sim 4200$ and $\sim 2500$ planetesimals,
for mass enhancement factors of 5 and 3, respectively. We limit our selection of the 
$\alpha$ and $\beta$ parameter space to models for which outward migration due to 
corotation torques is possible (the conditions for this can be determined by requiring
the entropy-related and vorticity-related horseshoe drag terms in Sect.~\ref{noniso}
to exceed the Lindblad torques). Our simulation parameters are detailed in 
Table~\ref{simparam}.

We set an inner edge to our simulations at 1 AU, and any body that migrates
inside this boundary, such that its semimajor axis is less than 1 AU, is removed
from the simulation. Information, however, is stored about each body
as it crosses this boundary, allowing us to follow the longer term trajectories of 
individual planets to determine their final stopping location as the gas disc
disperses. This procedure is referred to as `single-body analysis' later in the
paper.

  \begin{table}
     \centering
      \caption{Simulation parameters.\label{simparam}}
      \begin{tabular}{@{}llllll@{}}
        \hline 
        Simulation & \fenh  & $\alpha$ & $\beta$ & $M_{\rm solid}$ & $a_{\rm snow}$ \\
      
        \hline

        M01A, M01B & 5 & 0.5 & 0.75 & 173 & 1.95 \\
        M02A, M02B & 3 & 0.5 & 0.75 & 104 & 1.95 \\
        M03A, M03B & 5 & 0.5 & 1 & 173 & 1.65 \\
        M04A, M04B & 3 & 0.5 & 1 & 104 & 1.65 \\
        M05A, M05B & 5 & 0.5 & 1.25 & 173 & 1.49 \\
        M06A, M06B & 3 & 0.5 & 1.25 & 104 & 1.49 \\
        M07A, M07B & 5 & 0.5 & 1.5 & 173 & 1.39 \\
        M08A, M08B & 3 & 0.5 & 1.5 & 104 & 1.39 \\
        M09A, M09B & 5 & 0.75 & 1 & 173 & 1.65 \\
        M10A, M10B & 3 & 0.75 & 1 & 104 & 1.65 \\
        M11A, M11B & 5 & 0.75 & 1.25 & 173 & 1.49 \\
        M12A, M12B & 3 & 0.75 & 1.25 & 104 & 1.49 \\
        M13A, M13B & 5 & 0.75 & 1.5 & 174 & 1.39 \\
        M14A, M14B & 3 & 0.75 & 1.5 & 104 & 1.39 \\
        M15A, M15B & 5 & 1 & 1.25 & 170 & 1.49 \\
        M16A, M16B & 3 & 1 & 1.25 & 107 & 1.49 \\
        M17A, M17B & 5 & 1 & 1.5 & 170 & 1.39 \\
        M18A, M18B & 3 & 1 & 1.5 & 107 & 1.39 \\
        M19A, M19B & 5 & 1.25 & 1.5 & 173 & 1.39 \\
        M20A, M20B & 3 & 1.25 & 1.5 & 104 & 1.39 \\
        R01A, R01B & 2 & 0.5 & 1.25 & 69.6 & 1.49 \\
        R02A, R02B & 1 & 0.5 & 1.25 & 36.6 & 1.49 \\
        R03A, R03B & 2 & 0.5 & 1.5 & 69.6 & 1.39 \\
        R04A, R04B & 1 & 0.5 & 1.5 & 36.6 & 1.39 \\
        R05A, R05B & 2 & 0.75 & 1.25 & 69.6 & 1.49 \\
        R06A, R06B & 1 & 0.75 & 1.25 & 36.1 & 1.49 \\
        R07A, R07B & 2 & 0.75 & 1.5 & 69.6 & 1.39 \\
        R08A, R08B & 1 & 0.75 & 1.5 & 36.1 & 1.39 \\
        E01A, E01B & 5 & 0.5 & 1.25 & 173 & 1.49 \\
        E02A, E02B & 3 & 0.5 & 1.25 & 104 & 1.49 \\
        
        \hline
      \end{tabular}
  \end{table}
  
  \begin{figure*}
    \begin{center}
      \includegraphics[width=180mm, clip]{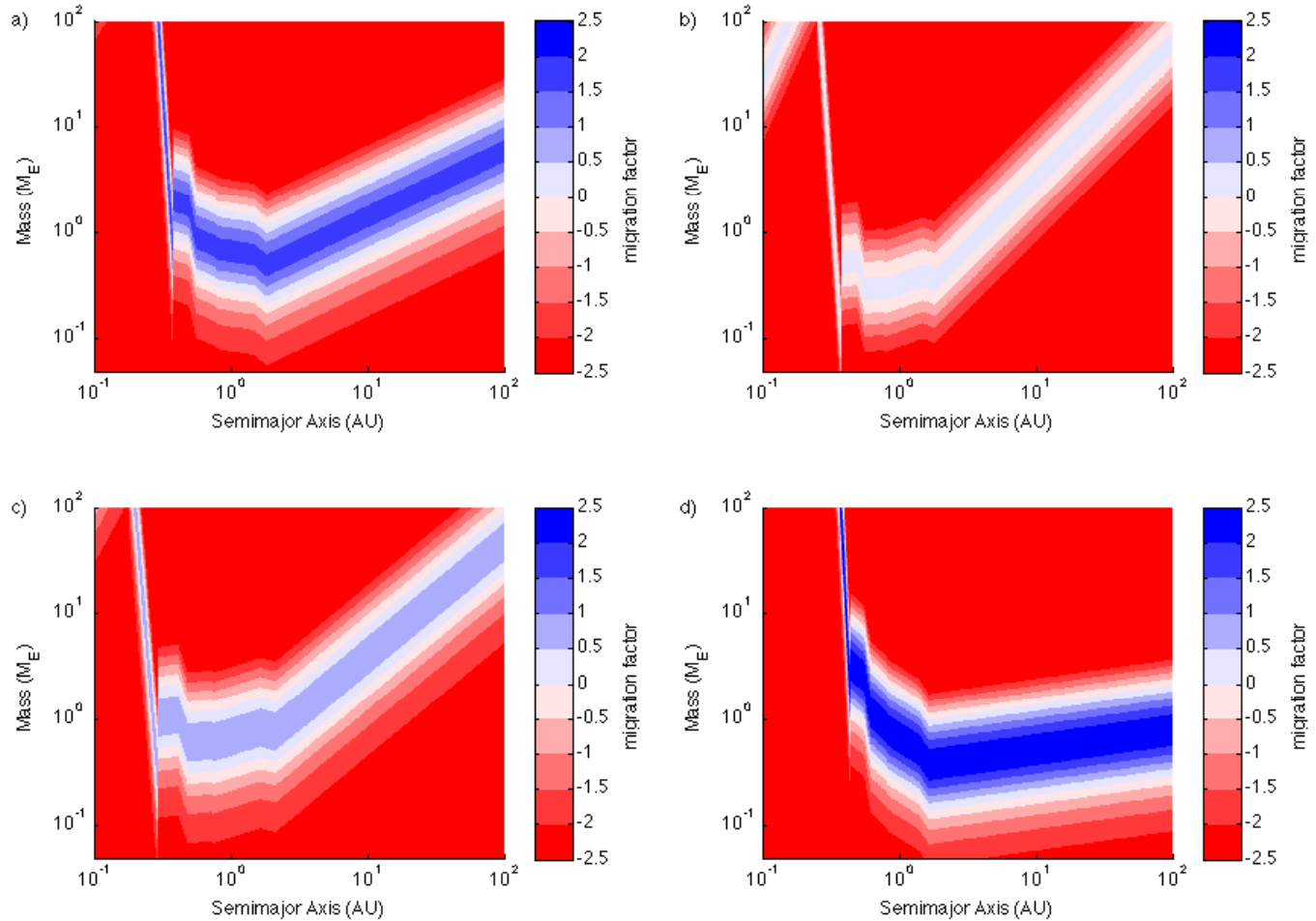}
      \caption{Contour plots showing regions of outward and inward migration in
the mass--semimajor axis plane for runs M05A (a), M16A (b), M03B (c) and M07B (d).
\label{plot:nonisocontours} }
    \end{center}
  \end{figure*}
  
  \section{Results}
  \label{endresults}
In this section we begin by describing common behaviour seen in many of the similations.
We introduce and discuss the concept of zero-migration lines, and their role in creating 
convergent migration within a swarm of growing protoplanets. We also discuss the coupling 
between the mass growth of protoplanets and their migration, and how rapid accretion by 
protoplanets can lead to migration into the central star.
  
We then describe the detailed evolution of a selection of individual runs (four runs in total),
followed by a summary of results across all of the simulations.  This includes the results of 
single-body analyses, where we investigate the evolution of bodies lost beyond the inner edge 
of the simulations (these are treated as isolated bodies, and so the analyses are limited in
their ability to provide accurate predictions about the nature of short-period systems).

Finally, we discuss briefly the effects of protoplanet eccentricity on the collective evolution
of the system, and present results in which the strength of corotation torques is attenuated
when a planet's eccentric orbit induces a radial excursion that is larger than the
horseshoe width.

Throughout this section, we define a gas giant as being a planet that has undergone runaway gas 
accretion, i.e. the sharp increase in mass shown in Fig.~\ref{plot:massgrowth}. This 
corresponds to a mass of approximately 30 M$_{\oplus}$.

\subsection{Common behaviour}
\label{commonbehaviour}

\subsubsection{Migration lines}
\label{miglines}
Consider a planet orbiting in a protoplanetary disc with power-law  
surface density and temperature profiles. If the planet sits
in the inner regions of the disc with high surface density and opacity,
such that the horseshoe libration time is significantly
{\it shorter} than the thermal/viscous
diffusion time across the horseshoe region, then the corotation torques
will saturate and be inoperable. The planet will migrate inward
rapidly as its evolution will be determined by Lindblad torques.

\begin{figure}
    \includegraphics[width=80mm, clip]{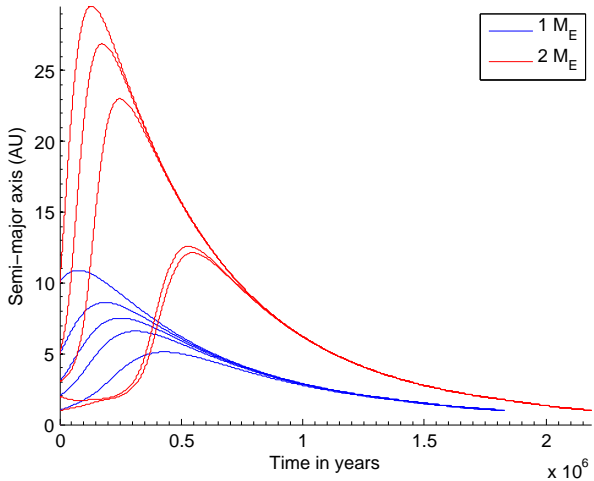}
    \caption{Migration lines showing convergent behaviour for 1 and 2 $M_E$ planets in a disc with initial conditions as in M05A.
      \label{plot:migrationlineM05A}}
  \end{figure}

  \begin{figure}
    \includegraphics[width=80mm, clip]{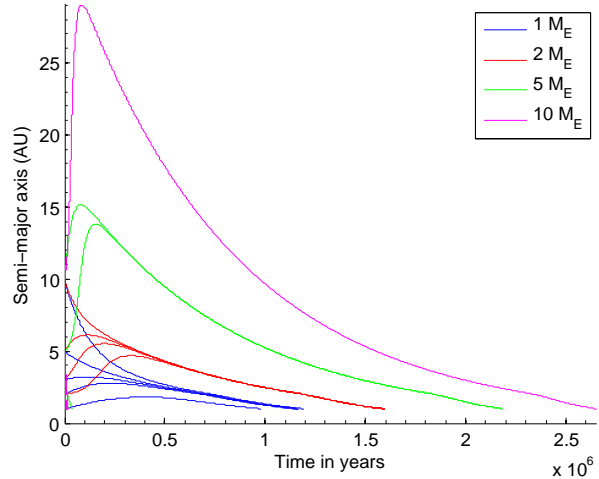}
    \caption{Migration lines showing convergent behaviour for planets with varying mass
 in a disc with initial conditions as in M03B.
      \label{plot:migrationlineM03B}}
  \end{figure}

Consider the same planet orbiting much further out in the disc
where the surface density and opacity are substantially reduced, 
such that the horseshoe libration time is much {\it longer} than the
thermal/viscous diffusion time. The disc-planet system is now close
to the locally isothermal limit, such that corotation torques will
be close to their linear value (see Sect.~\ref{noniso}). The migration will
again be inward because of the dominance of the Lindblad torques, but at
a reduced rate because of the contribution of positive corotation torques.

There exists an intermediate radial location in the disc where the
surface density and opacity allow the thermal/viscous diffusion
time to be approximately equal to the horseshoe libration time.
The corotation torque (horseshoe drag) will be close to its
maximum value here, and will possibly drive strong outward migration 
of the planet if the entropy gradient in the disc is steep enough.
As the planet migrates outward, however, the local disc surface
density and opacity decrease, decreasing the thermal diffusion time,
and reducing the efficacy of the positive corotation torque. 
Eventually the planet reaches a location
where the corotation and Lindblad torques exactly cancel, such that
the planet stops migrating. We refer to this location as the 
zero-migration line, and these are stable positions in the disc
for planets to reside. 

Given that the horseshoe libration time is shorter for more massive
planets, the zero-migration lines of heavier planets are located further
out in the disc where the thermal diffusion times are shorter. 
Heavier planets that form in inner disc regions will need to migrate out 
past lower mass bodies to reach their zero-migration lines, leading to 
convergent migration for protoplanets with different masses. 
Furthermore, protoplanets with the same mass try to migrate to the same
location in the disc. In principle, this should increase the rate of collisional 
planetary growth.

\begin{figure*}
    \begin{center}
      \includegraphics[width=180mm, clip]{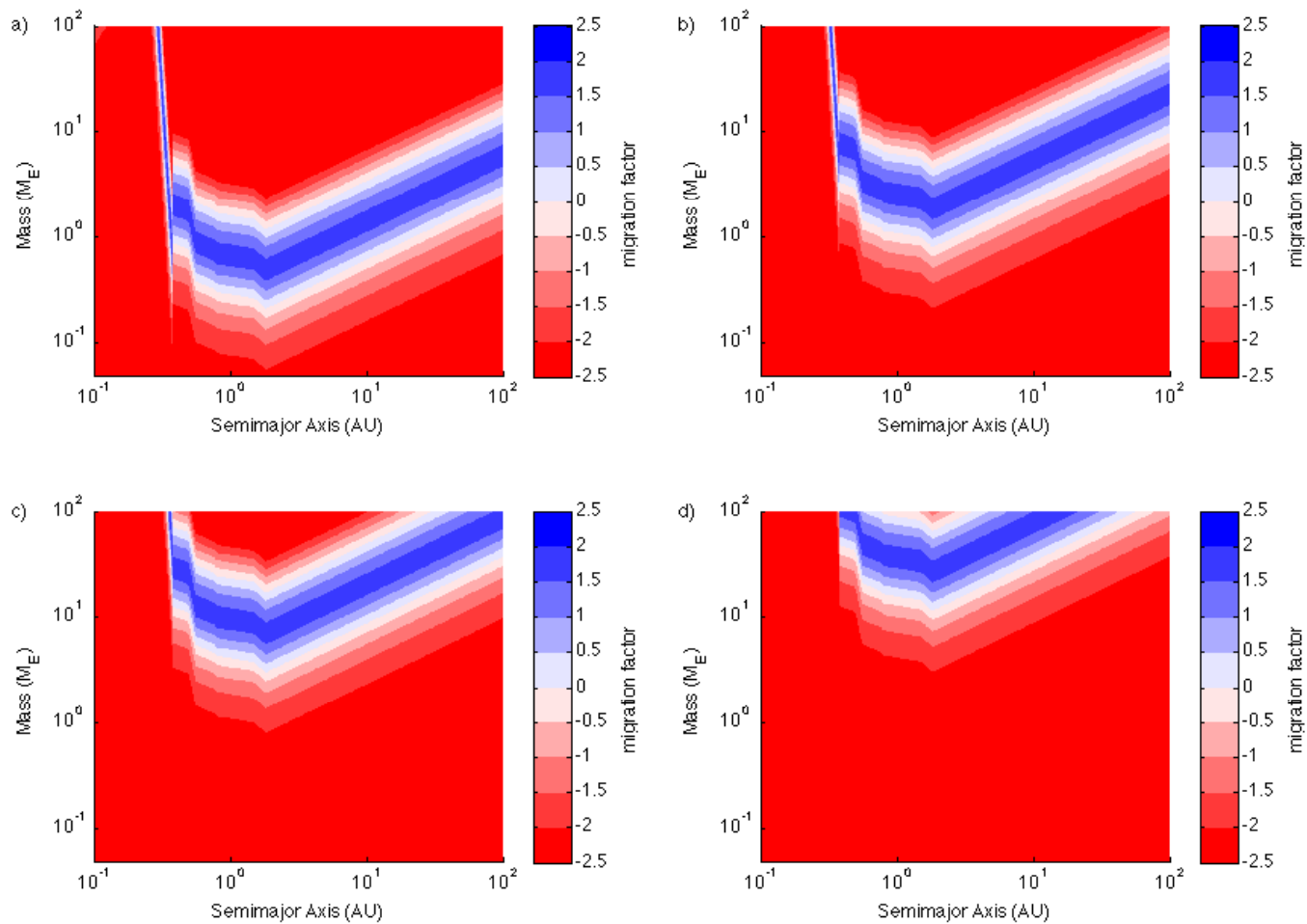}
      \caption{Contour plots showing regions of outward and inward migration for run M05A at t=0 years(a), t=1,000,000 years (b), t=2,000,000 years (c) and t=3,000,000 years (d).\label{plot:nonisocontours2} }
    \end{center}
  \end{figure*}
  
The behaviour described above is illustrated in the contour plots
shown in Fig.~\ref{plot:nonisocontours}, which display the 
value of the total torque in units of $\Gamma_0/\gamma$ (defined in 
Eq.~\ref{eq:Gamma0}), as a function of planetary mass and orbital position. 
The four panels correspond to the initial models M05A, M16A, M03B and M07B 
that are described in Table.~1. Regions coloured red correspond to 
strong inward migration (migration dominated by Lindblad torques). Regions 
coloured dark blue correspond to strong outward migration, and lightly coloured 
regions correspond to slow or zero migration. In general, rapid outward
migration is favoured in discs with relatively flat surface density 
profiles and steep temperature profiles.

In a steady-state disc, a planet of fixed mass that migrates
to its zero-migration line should stay there. Disc dispersal, however,
leads to a locally reducing surface density and opacity,
progressively shifting the zero-migration line inward. Consequently,
as the disc disperses, a planet sitting on a zero-migration line
drifts inward on the gas disc dispersal time scale. This behaviour is
shown in Figs.~\ref{plot:migrationlineM05A} and \ref{plot:migrationlineM03B},
which show the migration trajectories of planets of different mass
in the two disc models M05A and M03B that are dispersing on time scales
of 1 Myr (similar migration trajectories are shown in 
\citet{LyraPaardekooper2010}). Planets of a given mass starting at 
different locations
tend to migrate outward and eventually join the same migration line,
which then moves inward as the disc disperses. The behaviour of the
contours shown in the top left panel of Fig.~\ref{plot:nonisocontours}
under the action of disc dispersal are shown in 
Fig.~\ref{plot:nonisocontours2}. It is clear that as the disc
becomes increasingly optically thin, only heavy planets can sustain
outward migration (unless they become too massive and enter the type II
migration limit because of gap formation).

\begin{figure*}
    \begin{center}
      \includegraphics[width=140mm, clip]{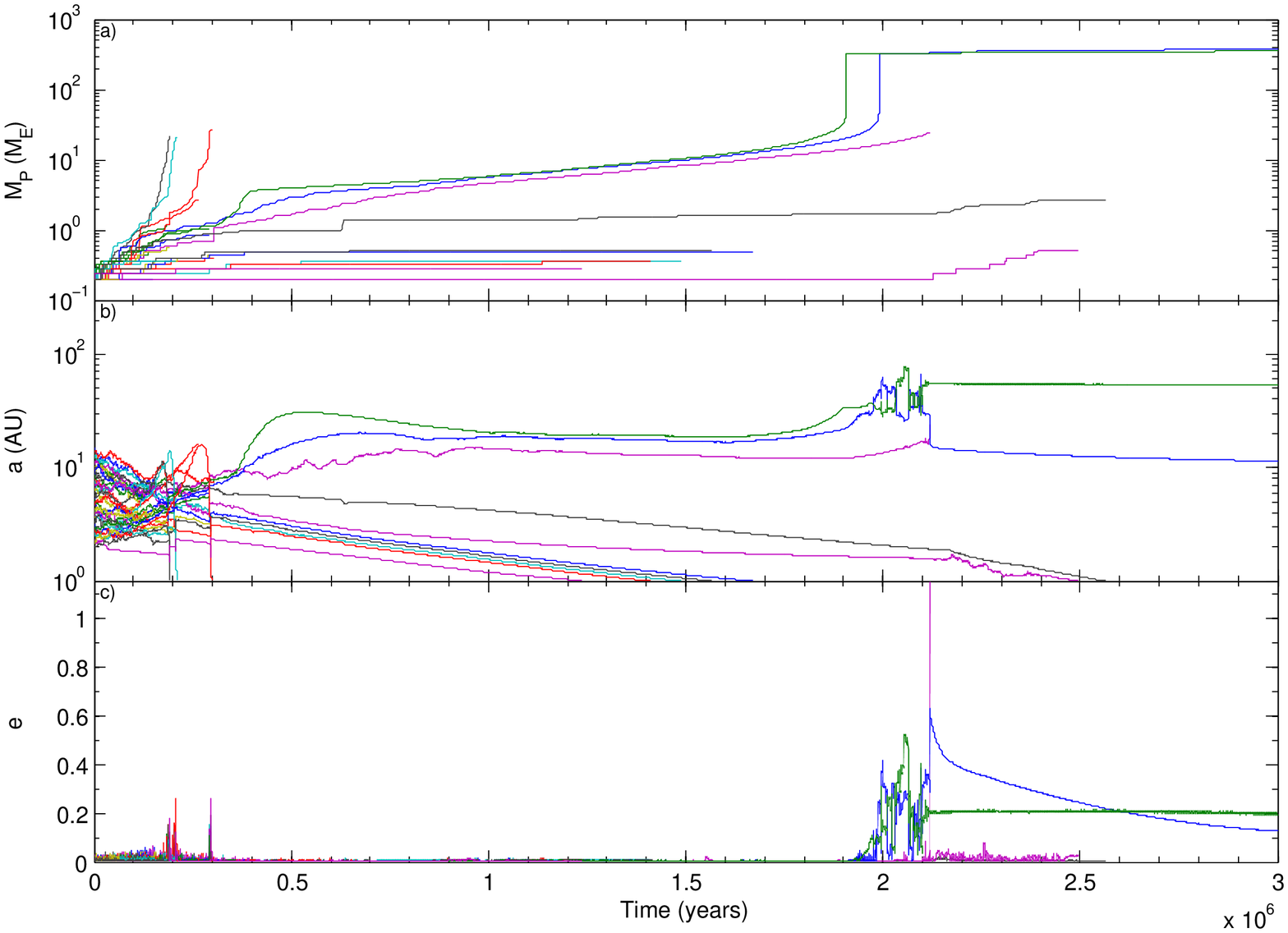}
      \caption{Evolution of the masses, semimajor axes and eccentricites
of all protoplanets simulation M05A.\label{plot:M05A_summary} }
    \end{center}
\end{figure*}

\subsubsection{Influence of mass growth}
\label{rapidgrowth}

A planet undergoing mass growth while sitting on a zero-migration 
line should migrate outward as it accretes, since zero-migration
lines for heavier planets lie at larger distances from the central star.
However, if the mass growth rate of the planet exceeds the outward 
migration rate due to corotation torques, then a very different
fate awaits it. Consider a planet of mass $\sim 0.5$ M$_{\oplus}$
sitting at $\sim 2$ AU in the top left panel of 
Fig.~\ref{plot:nonisocontours2}. Rapid growth of the planet 
up to 10 M$_{\oplus}$ in less than 1 Myr will put the planet
in the regime of rapid inward migration, as its trajectory in
the figure will be almost vertical, moving it out of the blue region
and into the red one. Very rapid growth of planets, therefore, may
not lead to strong outward migration but instead may cause planets
to migrate rapidly into the central star. The timing of the growth
of planets is crucial in determining their long-term survival.
      
  \subsection{Individual runs}
  \label{indruns}
  
  \subsubsection{Run M05A}
  \label{runM05A} 
Run M05A has an initial disc mass equivalent to 5 times the MMSN,
$\alpha=0.5$ and $\beta=1.25$. The magnitude and sign of
migration torques  $(t=0)$
are shown in the top left panel of Fig.~\ref{plot:nonisocontours},
and the effects of disc dispersal are demonstrated in 
Fig.~\ref{plot:nonisocontours2}. It is clear that planets with
masses in the range $0.2 \le m_{\rm p} \le 1$ M$_{\oplus}$, initially
located in the disc region 1 -- 10 AU, can undergo strong
outward migration. Growth of planets to masses of a few M$_{\oplus}$
may lead to outward migration over distances of tens of AU.

The time evolution of planet masses (top panel), semimajor axes (middle panel)
and eccentricities (bottom panel) are shown in Fig.~\ref{plot:M05A_summary}.
During the first 0.3 Myr, we observe that three planets grow in mass
rapidly, and undergo outward migration to $\sim 10$ AU. The mass growth
occurs as a result of planetesimal accretion and planet-planet inelastic
collisions, and the rapid growth is assisted by convergent migration
within the protoplanet swarm and by the gas envelopes that form within the
planet Hill spheres. When the planet masses exceed $\sim 20$ M$_{\oplus}$,
however, their migration direction changes and they undergo very rapid
inward migration through the planetary swarm and interior to 1 AU,
the inner boundary of the simulation. During the rapid inward migration, there
is very little accretion by these bodies, but they temporarily excite
the eccentricities of the other bodies in the system (see the bottom panel
of Fig.~\ref{plot:M05A_summary} between 0.2 -- 0.3 Myr).

Between the times 0.3 -- 0.5 Myr, we observe that three bodies grow to
masses larger than 1 M$_{\oplus}$. The largest of these grows to 
$\sim 4$ M$_{\oplus}$ and migrates outward rapidly to $\sim 30$ AU,
the location of its zero-migration line. We refer to this as ``planet A''.
A second planet (``planet B'') grows to a mass $\sim 3$ M$_{\oplus}$
by 0.5 Myr, and migrates out to its zero-migration line at $\sim 20$ AU.
A third planet (``planet C'') reaches a mass of $\sim 2$ M$_{\oplus}$
at 0.5 Myr and migrates out to 10 AU. Although the protoplanet/planetesimal
disc of solids is truncated at 15 AU in the initial simulation set-up,
outward migration of planets and gravitational scattering transports
planetesimals into the outer disc where they are accreted by planets
A and B (planet C continues to reside within the original planetesimal disc). 
Gas accretion ensues once these bodies exceed
3 M$_{\oplus}$, and we see their masses grow smoothly up to
20 -- 30 M$_{\oplus}$ between the times 0.5 -- 2 Myr. During this time
the zero-migration lines move inward (observe the modest inward migration
in the middle panel of Fig.~\ref{plot:M05A_summary} between 0.5 -- 2 Myr),
but continued mass growth helps to counterbalance this effect, and
prevents substantial inward migration. At approximately 2 Myr,
planets A and B undergo rapid gas accretion and grow to become Jovian-mass
giant planets (further gas accretion occurs at the viscous-supply rate).
The rapid mass growth induces dynamical instability between planets A and B,
causing them to undergo a period of gravitational scattering and eccentricty
growth (bottom panel of Fig.~\ref{plot:M05A_summary}). The scattering
eventually causes planets B and C to collide at 2.1 Myr (when planet C has a 
mass of 24 M$_{\oplus}$), leaving two giant planets on eccentric, non-crossing
orbits with semimajor axes of 12 and 55 AU. The inner planet has a total mass
of 374 M$_{\oplus}$, and a solid core mass of 27.6 M$_{\oplus}$, at the
end of the simulation. The outer planet has a total mass of 352 M$_{\oplus}$,
and a solid core mass of 10.1 M$_{\oplus}$.

During the formation of the outer gas giant planets, between time 0.5 -- 2.5 Myr,
only modest planetary growth occurs in the inner system.
An inner resonant convoy, similar to those discussed by 
\cite{McNeilDuncan2005} and \cite{CresswellNelson} migrates interior to
1 AU by $\sim 1.6$ Myr, driven by a more rapidly migrating 0.5 M$_{\oplus}$
body. This leaves behind two planets that grow to become $\sim 3$ and 0.4 M$_{\oplus}$
before migrating interior to 1 AU at $\sim 2.5$ Myr.

  \subsubsection{Run M16A}
  \label{runM16A} 
Run M16A has a disc mass equivalent to 3 times the MMSN, $\alpha=1$ and $\beta=1.25$.
The migration behaviour at $t=0$ is illustrated by the contours displayed in the
top right panel of Fig.~\ref{plot:nonisocontours}, and it is clear that 
outward migration is considerably weaker in this model than in the previously
described run M05A. Furthermore, the steepness of the outward migration
region as one moves to higher planet masses indicates that the radial
extent of outward migration is also reduced relative to model M05A. 
Placed in the initial disc model, a planet with $m_{\rm p} < 1$ M$_{\oplus}$ 
orbiting at 1 AU
will not undergo outward migration at all, but will instead migrate inward only.
Rapid planetary growth is therefore expected to result in much of the solid
disc material migrating in toward the star.

\begin{figure*}
    \begin{center}
      \includegraphics[width=140mm, clip]{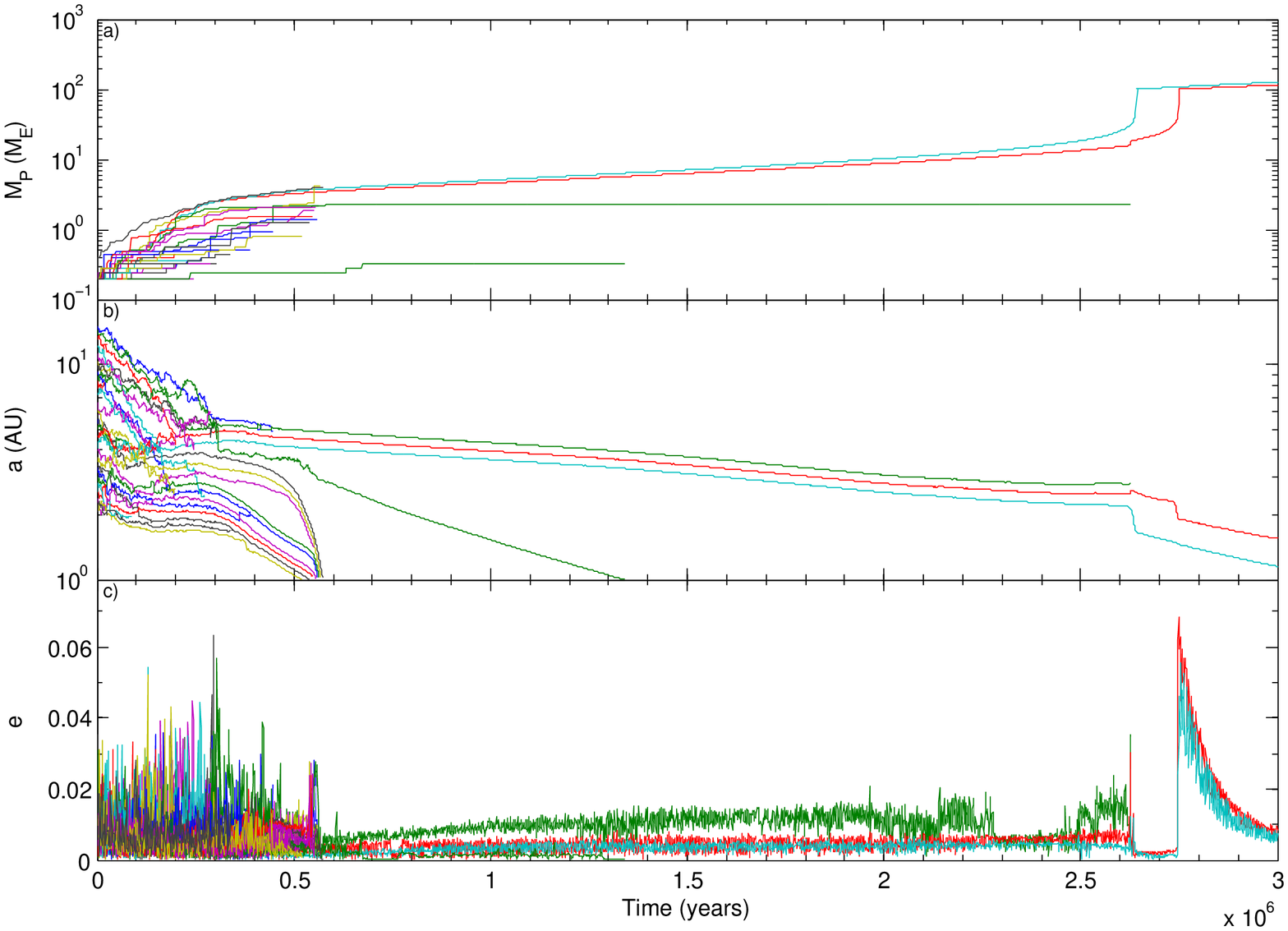}
      \caption{Evolution of the masses, semimajor axies and eccentricities
of all protoplanets in simulation M16A.\label{plot:M16A_summary} }
    \end{center}
  \end{figure*}

The evolution of the planetary masses (upper panel), semimajor axes
(middle panel) and eccentricities (bottom panel) are shown in 
Fig.~\ref{plot:M16A_summary}. Protoplanets located initially beyond $\sim 2$ AU
in this disc with masses $\simeq 0.02$ M$_{\oplus}$ (the inital masses of
protoplanets in the model) are expected to migrate inward, and looking
at the middle panel of Fig.~\ref{plot:M16A_summary} we see obvious evidence
for this migration occuring within the first 0.3 Myr. Looking at the inner part
of the system during the first 0.5 Myr, we observe two examples of 
resonant, inward migrating convoys being established. The first to form
consists of the six innermost protoplanets in the system. All masses of
these planets are $< 1$ M$_{\oplus}$, except for the outermost body, whose
mass has grown to $\sim 1$ M$_{\oplus}$. The more rapid migration of this
body drives the inward migration of the whole convoy. At a time of $\sim 0.4$
Myr, we see that the next three nearest protoplanets to the central star 
begin to undergo rapid inward migration, and this is driven by the formation
of a $\sim 5$ M$_{\oplus}$ body who's progenitor protoplanet was located at
$\sim 4$ AU. The growth of this protoplanet induces rapid inward migration,
with the system of inner planets being swept interior to 1 AU at $t=0.55$ Myr.

Three planets initially located at $\sim 5$ AU become physically detached
from the rest of the system after $\sim 0.5$ Myr, as shown in the middle panel of
Fig.~\ref{plot:M16A_summary}. These bodies have all grown to masses between
2 -- 5 M$_{\oplus}$ within this time. The outermost $\sim 2$ M$_{\oplus}$ 
body becomes isolated from planetesimals in the disc such that its
mass does not grow after 0.5 Myr. This isolation occurs in large part
because the two more massive neighbouring planets accrete the nearby planetesimals.
Having achieved masses in excess of 3 M$_{\oplus}$, these two planets are able to accrete
gas. As they do so, they sit on their zero-migration lines and undergo slow
inward migration, where the speed of migration is attenuated by the continuing
gas accretion and mass growth (the planets try to migrate outward to
the zero-migration lines for more massive planets as they grow, at the same time as the
zero-migration lines move inward as the gas disc evolves). After $\sim 2.6$ Myr,
the innermost planet reaches a mass of $\sim 30$ M$_{\oplus}$ and undergoes rapid
gas accretion to become a Saturn-mass gas giant. The rapid mass growth dynamically
disturbs the system, as observed in the middle and bottom panels
of Fig.~\ref{plot:M16A_summary}, causing the outer 2 M$_{\oplus}$ planet to
collide with the middle planet. Shortly after, this merged  planet undergoes
rapid gas accretion to also become a Saturn-mass gas giant. Saturnian rather Jovian
masses are achieved because accretion occurs late in the disc lifetime, such that
the gas isolation mass limits the envelope mass to $\sim 100$ M$_{\oplus}$.

At the end of the simulation we have an inner planet of total mass 115 M$_{\oplus}$,
with solid core mass 11 M$_{\oplus}$, orbiting at 2.3 AU, and an outer planet
with total mass 127 M$_{\oplus}$, solid core mass 8.8 M$_{\oplus}$, orbiting at
3.1 AU.

  \subsubsection{Run M03B}
  \label{runM03B}
Run M03B has a disc mass equivalent to 5 times that of the MMSN,
$\alpha=0.5$ and $\beta=1$. The migration behaviour of this
model at $t=0$ is illustrated by the lower left panel in Fig.~\ref{plot:nonisocontours},
showing that this disc is intermediate between the two previous models discussed 
(M05A and M16A) in terms of the strength of outward migration.

\begin{figure*}
    \begin{center}
      \includegraphics[width=140mm, clip]{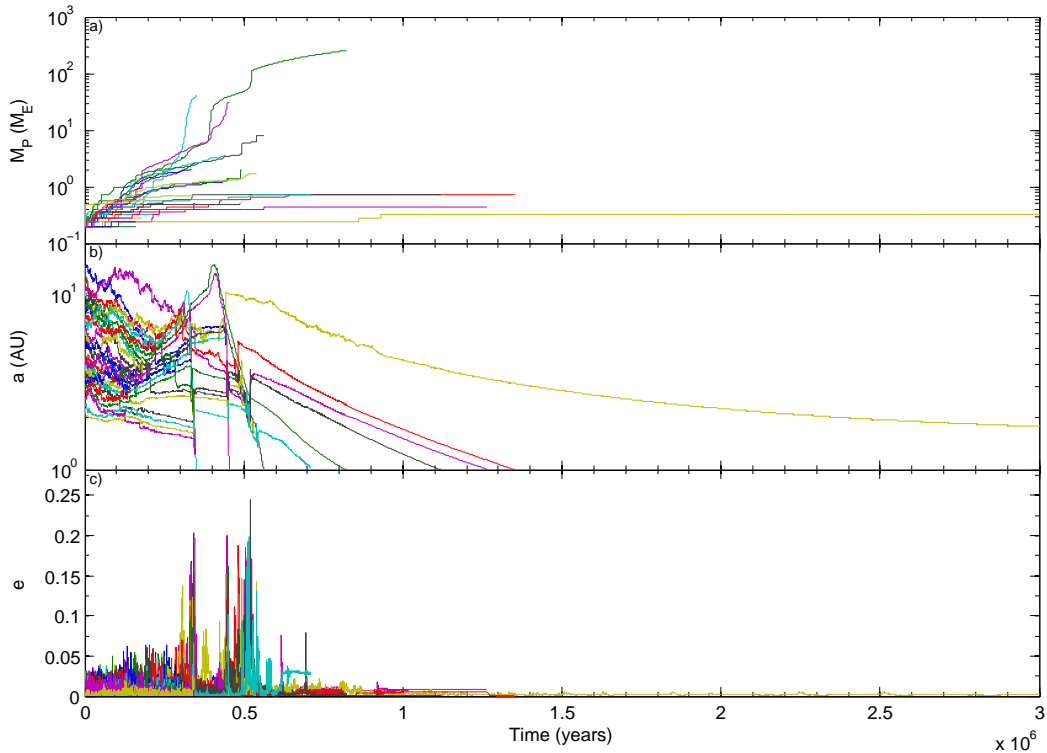}
      \caption{Evolution of the masses, semimajor axes and eccentricities
 of all protoplanets in simulation M03B.\label{plot:M03B_summary} }
    \end{center}
  \end{figure*}

The evolution of the planetary masses (upper panel), semimajor axes (middle panel)
and eccentricities (bottom panel) are shown in Fig.~\ref{plot:M03B_summary}. 
As expected from comparing the migration behaviour of the runs M16A and M03B in
Fig.~\ref{plot:nonisocontours}, the initial stages of run M03B show some
similarities to run M16A. Protoplanets initially located 
in the region of the protoplanet/planetesimal disc between 10 -- 15 AU
migrate inward to the region centred around 2 -- 3 AU.
The inner planets, however, do not show a strong tendency to migrate
inward (differing in this regard from run M16A), and the convergent
migration stimulates substantial growth within the protoplanet swarm,
as seen in the upper panel of Fig.~\ref{plot:M03B_summary}, where
the planet masses are seen to increase during the first 0.5 Myr,
and in the middle panel where it is clear that collisional growth
reduces the number of planets. The strong planetary growth, however,
also leads to rapid inward migration. Bodies that reach masses in
excess of 20 M$_{\oplus}$ undergo rapid inward migration through 
the disc of protoplanets/planetesimals and interior to 1 AU,
exciting the eccentricities of the remaining planets as they do so.
The bodies that rapidly migrate through the inner boundary at
1 AU would hit the star if their long-term evolution were followed.

\begin{figure}
    \begin{center}
      \includegraphics[width=80mm, clip]{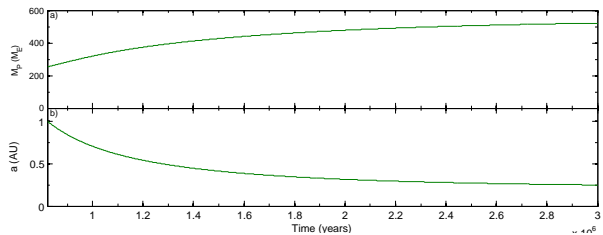}
      \caption{Evolution of mass and semimajor axis of single body extension run for the 
 hot Jupiter in simulation M03B starting at 0.821 Myr.\label{plot:M03B_singlebody} }
    \end{center}
  \end{figure}

One of the quickly migrating planets (shown by the upper green line in the top
panel of Fig.~\ref{plot:M03B_summary}) grows to be massive enough (approximately
30 M$_{\oplus}$ of solids) to undergo rapid gas accretion during the inward migration.
It reaches its gas isolation mass at a mass equal to 114 M$_{\oplus}$, and
transitions to type II migration, drifting interior to 1 AU shortly
after 0.8 Myr has elapsed. At this point in time, the planet mass is $\sim 250$ 
M$_{\oplus}$, and it is undergoing gas accretion from the disc at the 
viscous-supply rate. We have followed the evolution of this body after it has 
traversed the inner boundary of the simulation, treating it as an isolated body 
and neglecting its interaction with other bodies in the system (we refer to 
this as single-body analysis). The evolution is displayed in 
Fig.~\ref{plot:M03B_singlebody}, and we see that the planet reaches a semimajor axis of 
0.25 AU and has a mass of 524 M$_{\oplus}$ after 3 Myr, making it an excellent candidate for 
a hot Jupiter.

No other planets grow substantially during the evolution of this run.
Fig.~\ref{plot:M03B_summary} shows that only a single planet
with mass $\sim 0.35$ M$_{\oplus}$ survives beyond 1 AU, coming
to rest at a semimajor axis of $\sim 1.9$ AU.

    \subsubsection{Run M07B}
    \label{runM07B}
The final run we describe in detail is M07B, which has a mass equivalent
to 5 times the MMSN, $\alpha=0.5$ and $\beta=1.5$. The steep temperature
gradient and relatively shallow surface density gradient allow this
disc model to support strong outward migration over a large radial extent,
as illustrated by the contours in the bottom right panel of 
Fig.~\ref{plot:nonisocontours}.
This plot demonstrates clearly that sub-Earth mass bodies orbiting in the
vicinity of 1 AU will experience strong outward migration, possibily out
beyond 100 AU.

 \begin{figure*}
    \begin{center}
      \includegraphics[width=140mm, clip]{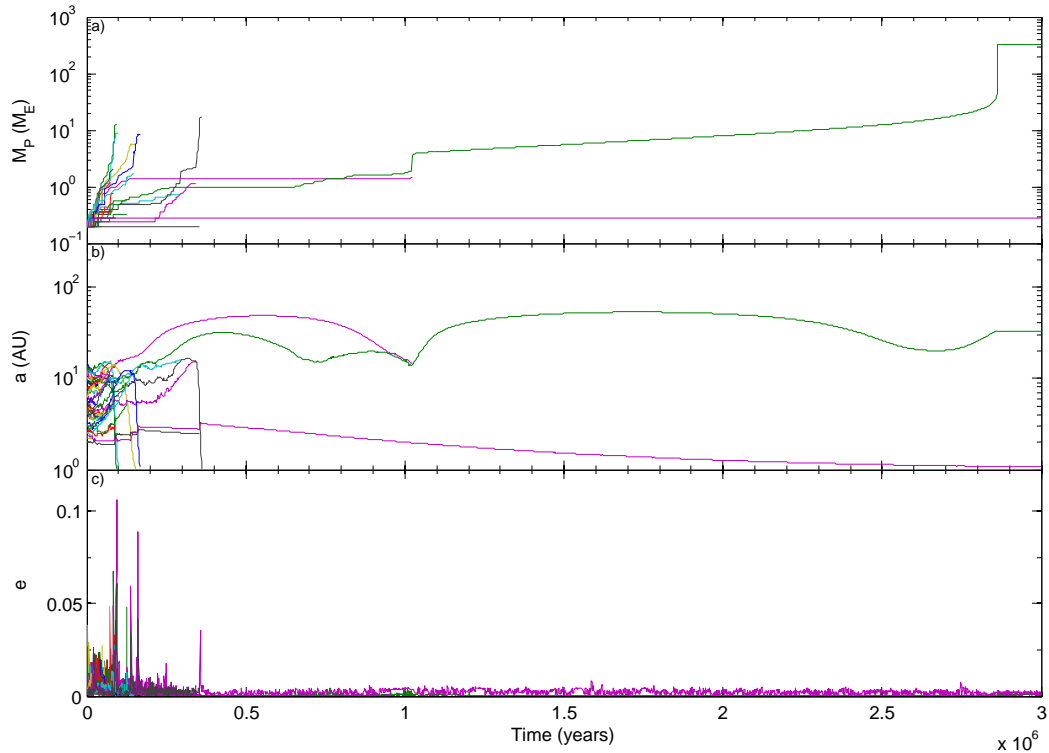}
      \caption{Evolution of masses, semimajor axes and eccentricities
 of all protoplanets in simulation M07B.\label{plot:M07B_summary} }
    \end{center}
  \end{figure*}

The evolution of planetary masses (top panel), semimajor axes (middle panel)
and eccentricities (bottom panel) are shown in Fig.~\ref{plot:M07B_summary}.
The initial growth and outward migration of protoplanets in the
inner region of the swarm leads to strongly convergent migration, and
rapid growth of a number of bodies up to $\sim 10$ M$_{\oplus}$ within
the first 0.3 Myr. As observed in the previously described runs, this
leads to rapid inward migration of these planets because the
horseshoe libration time becomes much shorter than the thermal/viscous
diffusion time for these bodies. After 0.5 Myr, there are only three bodies
left in the simulation: two planets orbiting at $\sim 15$ AU each with masses
$\sim 2$ M$_{\oplus}$; one protoplanet with mass $\sim 0.3$ M$_{\oplus}$
orbiting at $\sim 2$ AU. The two outer bodies collide shortly after 0.5 Myr,
and the resulting planet undergoes slow gas accretion, migrating outward as
it does so. After 2.8 Myr it undergoes rapid gas accretion and becomes
a Jovian mass (319 M$_{\oplus}$) gas giant planet, with a 5.2 M$_{\oplus}$ solid core,
orbiting at  $\sim 33$ AU.
As in model M05A, we find that gas giant planets can be formed at large
radius from the central star by the migration and gas accretion onto
a solid core. In both of these models (and others not discussed in detail),
the mode of formation is one in which an initial generation of massive 
protoplanets form and migrate in toward the central star, followed by
a second generation of more isolated lower mass cores that can migrate out 
slowly and accrete gas at the same time. Such a model may provide a natural
explanation for the massive planets orbiting at large
distance from their host stars, such as HR 8799 \citep{Marois}, Fomalhaut \citep{Kalas},
and Beta Pictoris \citep{Lagrange}, that are being discovered by direct imaging
surveys.

  \subsection{Summary of all runs}
  \label{allruns}
\begin{figure*}
  \begin{center}
  \includegraphics[width=140mm, clip]{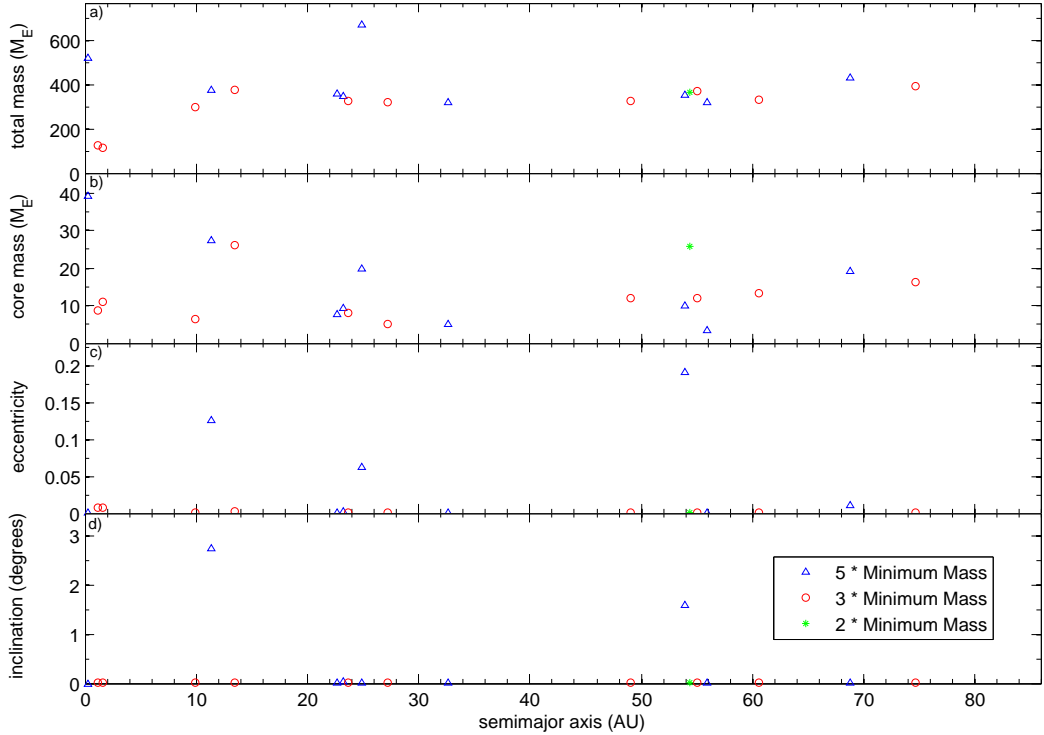}
  \caption{Summary of total masses, core masses, eccentricity and inclination against semimajor axis for all gas giant planets formed.\label{plot:totalsummary} }
  \end{center}
  \end{figure*}

We ran 40 simulations with $f_{\rm enh}=5$ or 3, surface density power-law indices
in the range $0.5 \le \alpha \le 1.25$, and temperature power-law indices satisfying
$0.75 \le \beta \le 1.5$. In total 19 gas giant planets were formed in these runs,
and their properties are summarised in Fig.~\ref{plot:totalsummary} and Table~\ref{results}.
The giant planet masses range from 115 to 670 M$_{\oplus}$, and have solid core masses
in the range 3.6 -- 39 M$_{\oplus}$. Looking at the upper panel of Fig.~\ref{plot:totalsummary},
we see that most giant planets are grouped within the mass range 320 to 400 M$_{\oplus}$,
and this is very likely to be an artifact of our gas accretion procedure that
limits the planet mass obtained during rapid gas accretion to be the Jovian mass. 
A more sophisticated procedure would be sensitive to local conditions in the disc,
and result in a broader spread of planet masses, and this is clearly
one future improvement that we will need to implement in our modelling procedure.
Nonetheless, we do also obtain giant planets outside of this mass range.
Run M03B formed a 523.8 M$_{\oplus}$ planet, as discussed in Sect.~\ref{runM03B},
due to a gas giant forming within the first 0.5 Myr, and subsequently 
accreting viscously and migrating via type II migration to its final location at 0.25 AU.
The heaviest planet formed during run M11A, and this was the result of two gas giant planets 
colliding, having each formed at between 20 and 30 AU from the central star due to their cores 
migrating outward. Employing a pure hit-and-stick prescription for planetary collisions, however, 
probably leads to an overestimate of the final mass of this planet.
Three planets were formed with masses below the imposed Jovian-mass limit.
Run M12B produced a 296 M$_{\oplus}$ planet orbiting at 9.8 AU, and as
described in Sect.~\ref{runM16A}, run M16A produced a pair of Saturn-mass objects
orbiting at 1.15 and 1.55 AU. These planets formed late in the disc lifetime,
such that their gas isolation masses were below the Jovian mass.
  
   \begin{figure*}
  \begin{center}
  \includegraphics[width=140mm, clip]{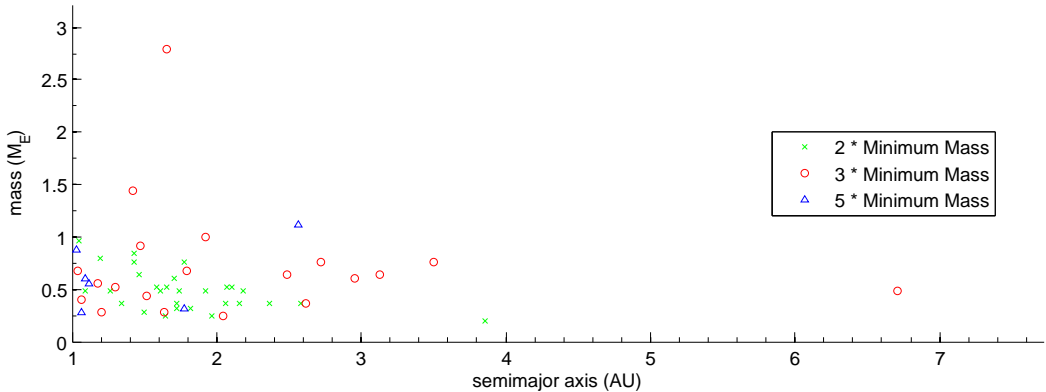}
  \caption{Summary of masses against semimajor axis for all small surviving 
planets outside 1 AU.\label{plot:smallerbodiessummary} }
  \end{center}
  \end{figure*}
  
Most of the giant planets formed at semimajor axes substantially  beyond 10 AU. Indeed, 
only 4 out of 19 giant planets formed interior to 10 AU. The reason for this is that
the most common mode of gas giant planet formation in the simulations was the formation
of a core of modest mass in the interior disc, that then migrates outward over large 
distances before accreting a gas envelope. 
Many massive cores formed during the early stages of the disc life times
in the simulations, and were able to undergo gas accretion.
Their rapid inward migration, however, prevented them from surviving.
Giant planets that form closer to the
star and survive tend to be in disc models (M03B and M16A) that generate weaker 
corotation torques.

There are three simulations that lead to the formation of surviving multiple giant 
planets. Run M05A produces a pair of Jovian mass planets orbiting at 11.4 and 53.9 AU,
as described in Sect.~\ref{runM05A}, and M05B also produced a pair of Jovian mass
objects orbiting at 23.3 and 68.8 AU. Note that these runs were identical apart from
the random number seed used to generate initial conditions. Run M16A produced a pair
of Saturn-mass planets orbiting at 1.15 and 1.55 AU. One consequence of this is that
almost all giant planets formed in the simulations are on circular, non-inclined orbits.
The only planets with significant eccentricities are those in run M05A, where 
gravitational scattering during the formation caused the growth of eccentricity.

One surprising result to come out of the simulations is the lack of
correlation between initial disc mass and the frequency of giant planet
formation. Discs with $f_{\rm enh}=5$ formed 9 giants and those with
$f_{\rm enh}=3$ formed 10. This led us to question whether less massive discs 
might be able to form gas giants. To examine this, we performed additional simulations 
(labelled `R' in Table \ref{simparam}) based on the most succesful models in the 
$f_{enh}$=3 and 5 runs. All barring one failed to produce any gas giants. 
Run R07B, a $f_{enh}$=2 disc, did produce a single 
Jovian mass gas giant (shown in Fig.~\ref{plot:totalsummary}).

In addition to the giant planets discussed above, the simulations also resulted in the
formation and survival of lower mass bodies beyond 1 AU in the disc. These are shown in
Fig.~\ref{plot:smallerbodiessummary}. The rapid growth of cores, followed by
rapid inward migration, has the effect of clearing much of the solid material
from beyond 1 AU, and the outward migration of modest sized cores that evolve into gas giants
also clears this region. Nonetheless, terrestrial mass bodies do form and survive in the 
simulations, although Fig.~\ref{plot:smallerbodiessummary} shows that these tend to
be in the lower mass discs. One noticeable and interesting observation about the simulation
results is the lack of super-Earth and Neptune mass planets. The rapid formation of massive
cores that undergo fast inward migration is a major cause of this (driven by efficient
capture of planetesimals and convergent migration), but a contributing factor is the
fact that planets with masses greater than 3 M$_{\oplus}$ can begin to undergo gas accretion
in our models.
A higher threshold for gas accretion would probably allow some of the giant planets that formed
to have maintained lower masses. These observations provide a useful guide to the types
of modifications that the modelling procedure requires in order to form planets with
the same characteristics as those which are contained in the extrasolar planet observational database.

   \begin{table*}
     \centering
     \begin{minipage}{160mm}
      \caption{Summary of gas giants formed.\label{results}}
      \begin{tabular}{@{}lllllllll@{}}
        \hline 
        Simulation & \fenh & $\alpha$ & $\beta$ & a (AU) & e & i (degrees) & mass ($\ME$) & core mass ($\ME$) \\
     
        \hline

        M03B & 5 & 0.5 & 1 & 0.24818 & 0.000001 & 0 & 523.79 & 39.39 \\
        M05A & 5 & 0.5 & 1.25 & 11.39435 & 0.125762 & 2.7366 & 374.36 & 27.57 \\
        M05A & 5 & 0.5 & 1.25 & 53.91049 & 0.191585 & 1.5949 & 352.2 & 10.11 \\
        M05B & 5 & 0.5 & 1.25 & 23.26593 & 0.00314 & 0.0299 & 351.47 & 9.51 \\
        M05B & 5 & 0.5 & 1.25 & 68.79704 & 0.011044 & 0.0195 & 433.61 & 19.18 \\
        M06A & 3 & 0.5 & 1.25 & 54.99131 & 0.000684 & 0.0026 & 369.38 & 12.11 \\
        M06B & 3 & 0.5 & 1.25 & 74.69739 & 0.000847 & 0.0012 & 392.5 & 16.1 \\
        M07A & 5 & 0.5 & 1.5 & 55.91612 & 0.000897 & 0.0006 & 319.86 & 3.6 \\
        M07B & 5 & 0.5 & 1.5 & 32.59897 & 0.00098 & 0.0057 & 319.24 & 5.19 \\
        M08A & 3 & 0.5 & 1.5 & 13.41661 & 0.001873 & 0.0106 & 374.61 & 26.13 \\
        M08B & 3 & 0.5 & 1.5 & 60.55699 & 0.000785 & 0.0021 & 333.38 & 13.35 \\
        M11A & 5 & 0.75 & 1.25 & 24.93238 & 0.063072 & 0.0053 & 669.88 & 19.7 \\
        M11B & 5 & 0.75 & 1.25 & 22.70169 & 0.000959 & 0.0069 & 361.13 & 7.79 \\
        M12B & 3 & 0.75 & 1.25 & 9.87598 & 0.000726 & 0.0022 & 296.43 & 6.47 \\
        M14A & 3 & 0.75 & 1.5 & 27.22201 & 0.000179 & 0.0046 & 323.17 & 5 \\
        M14B & 3 & 0.75 & 1.5 & 49.00695 & 0.000497 & 0.0036 & 328.24 & 12.07 \\
        M16A & 3 & 1 & 1.25 & 1.55415 & 0.008245 & 0.0008 & 114.91 & 10.9511 \\
        M16A & 3 & 1 & 1.25 & 1.15495 & 0.007372 & 0.0008 & 126.85 & 8.8343 \\
        M18B & 3 & 1 & 1.5 & 23.61206 & 0.000516 & 0.0111 & 325.91 & 7.95 \\
        R07B & 2 & 0.75 & 1.5 & 54.41991 & 0.000599 & 0.0016 & 367.3 & 25.77 \\
        
        \hline
      \end{tabular}
  \end{minipage}
  \end{table*}
   
  \subsubsection{Single-body analysis}
  \label{singlebody}
The inner edge of our simulations was set at 1 AU. We ran single body runs for each object
that was lost beyond this inner edge so as to identify any bodies that would have become short 
period gas giants, and to obtain an estimate of the distribution of smaller bodies in this 
inner region. These runs are effectively continuation runs, but with a smaller time step size, 
and a smaller inner boundary at 0.1 AU. It is important to note that these single body runs did 
not include the influence of any other planets in the system, and did not account for any material 
that would have been present between 0.1 -- 1 AU during the early evolution of the system. As such, 
the results merely provide a guide to the planets that can survive within this radial range.

\begin{figure*}
  \begin{center}
  \includegraphics[width=140mm, clip]{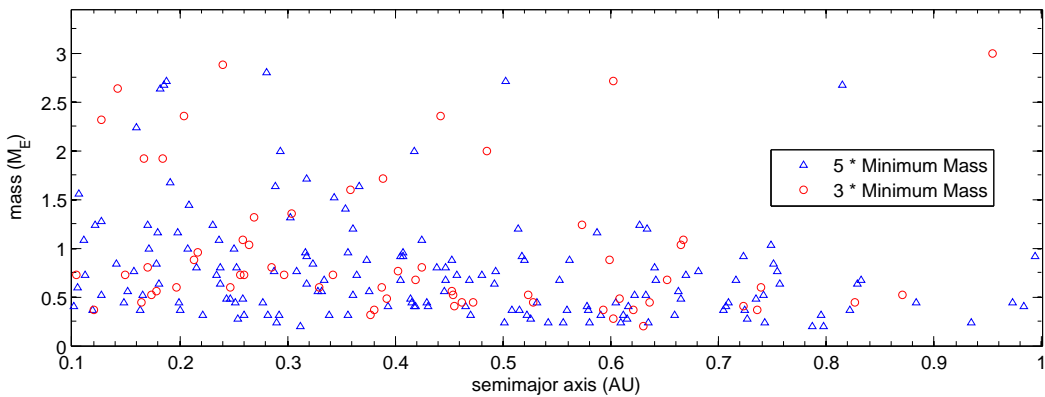}
  \caption{Summary of masses against semimajor axis for all small surviving planets interior to 1 AU.
 Note that these data were obtained using the single-body analysis described in the text.
 \label{plot:innerbodysurvived} }
  \end{center}
  \end{figure*}

Figure \ref{plot:innerbodysurvived} shows a summary of all non-giant planets left remaining in the 0.1 
to 1 AU region from all the $f_{enh}$=3 and 5 models. Objects with masses less than 1 M$_{\oplus}$ are 
clearly more common than those with larger masses because of their reduced migration rates.  Also, 
smaller semimajor axes seem the more likely outcome. All objects included in this figure have
masses below 3 M$_{\oplus}$ and have not been able to undergo gas accretion. The only gas giant
to survive in the region 0.1 -- 1 AU is the one described already in Sect.\ref{runM03B}.
 
\begin{figure*}
  \begin{center}
  \includegraphics[width=140mm, clip]{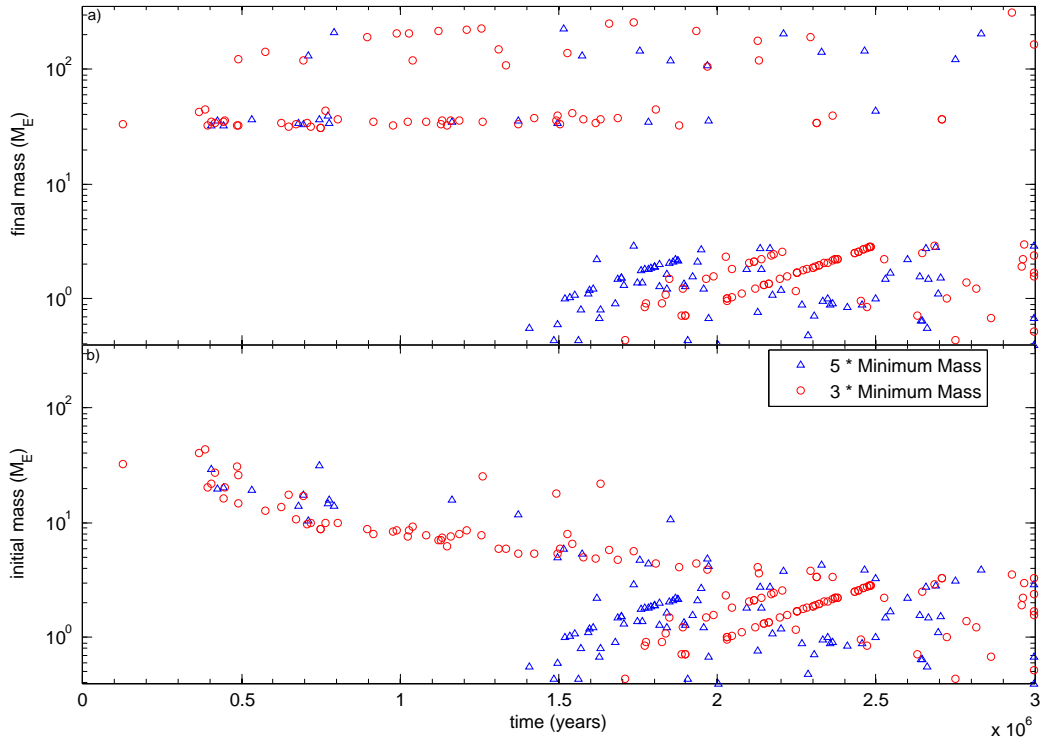}
  \caption{Summary of final masses and initial masses against time of loss for all planets that were lost beyond 0.1 AU.\label{plot:innerbodylost} }
  \end{center}
  \end{figure*}

A large number of bodies are lost beyond 0.1 AU, ranging from the 
smallest protoplanets all the way up to Jupiter sized gas giants, potentially providing the
central star with a significant enrichment of heavy elements. These bodies are summarised
in Fig.~\ref{plot:innerbodylost}, which shows the masses of the planets as they cross the
boundary at 1 AU in the lower panel, and their masses as they cross the boundary at 0.1 AU
in the upper panel. The horizontal axes show the time of loss through the boundary at 0.1 AU.
It is clear that a number of the massive cores that migrate through the 1 AU boundary
accrete gas and become gas giants, although their masses normally reach values between
100 -- 200 M$_{\oplus}$ because the gas isolation mass is below the Jovian-mass in
the inner disc.  Type II migration drives them through the boundary at 0.1 AU.
It is also clear that a number of bodies migrate inside 1 AU with masses in the range
4 -- 10 M$_{\oplus}$ and grow through gas accretion to masses between 30 -- 50 M$_{\oplus}$.
Type I migration, however, forces these bodies to migrate into the star before they can become
giants. Their corotation torques are saturated, and so rapid inward
migration is driven by Lindblad torques.

Some of the bodies passing through the 0.1 AU boundary at late times could have survived.
We ran extended single body runs for the five planets with masses greater than 25 M$_{\oplus}$
lost beyond 0.1 AU in the last 500,000 years of simulation time. Two collided with the central body at 
~2.75 Myr, but the other three survived at 0.086, 0.0428 and 0.016 AU with masses 344, 164 and 550 
M$_{\oplus}$, respectively. We have not included these results along with the other gas giants since 
their simulation conditions were overly simplified compared to the rest. All three bodies entered the 
1 AU zone with just a few Earth masses, and so would in reality have interacted with other bodies and 
planetesimals formed there which were not modelled. Also, the body ending up at 0.016 AU
would most likely have been accreted by the central star a short while later.

The survivability of these giant planets formed in single body analyses between 1 and 0.1 AU depends 
on exactly how the gas disc dissipates. The exponential dissipation of gas provides a reasonable 
approximation for the bulk of the gas dissipation when it is dominated by viscous
evolution \citep{FoggNelson2007}, but at later times the structure of a
viscously evolving disc that is being photoevaporated by UV radiation from the central
star changes substantially \citep{ClarkeSotomayor}  with a low density inner cavity being formed.
Clearly such a model needs to be incorporated into the simulation procedure outlined here
to make reasonable predictions about the nature of surviving short-period planets.
  
      \begin{figure*}
    \begin{center}
      \includegraphics[width=140mm, clip]{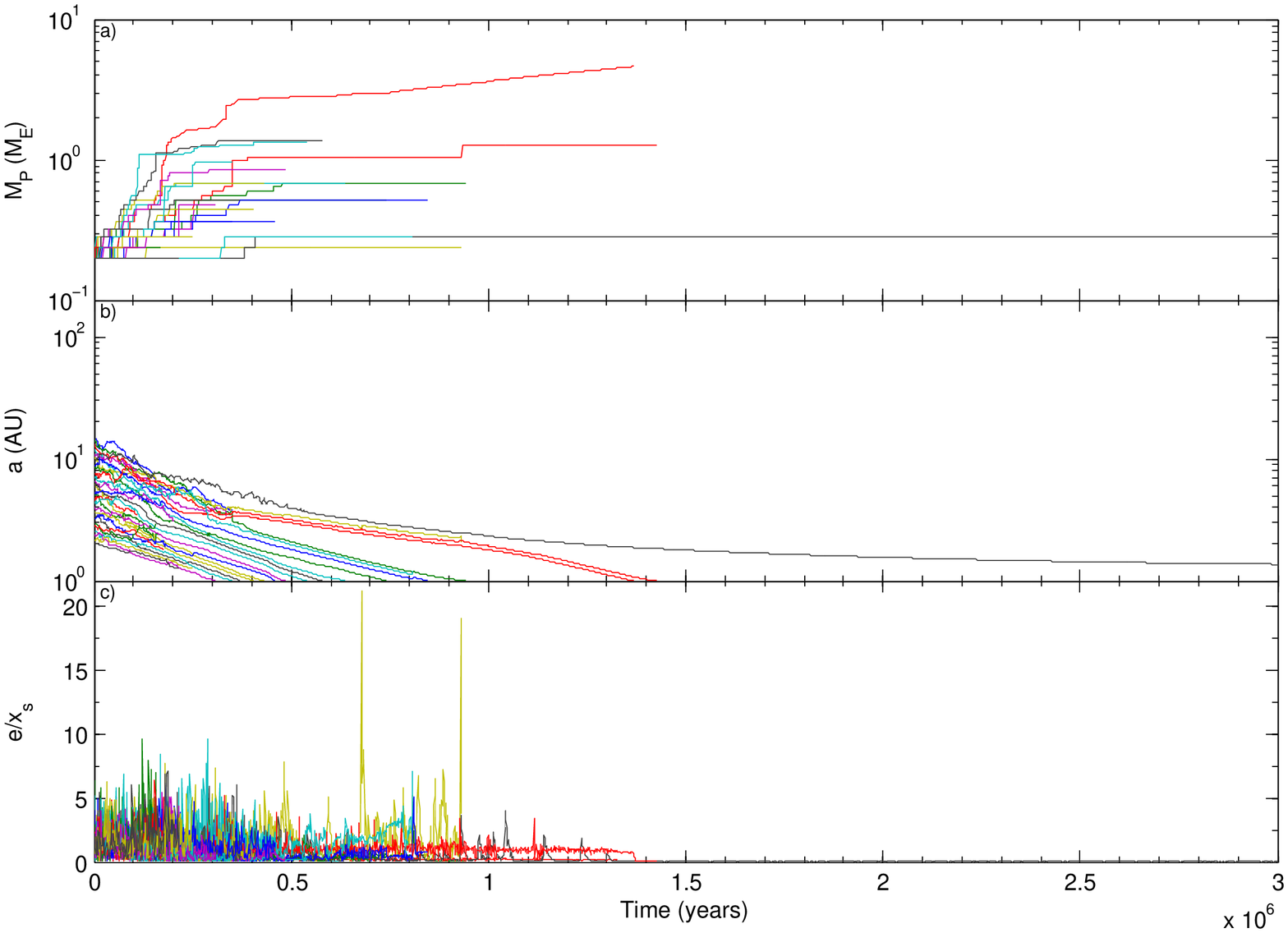}
      \caption{Evolution of masses, semimajor axes and eccentricities of all protoplanets 
in simulation E02B.\label{plot:E02B_summary_e_xs} }
    \end{center}
  \end{figure*}
  
    \begin{figure*}
    \begin{center}
      \includegraphics[width=140mm, clip]{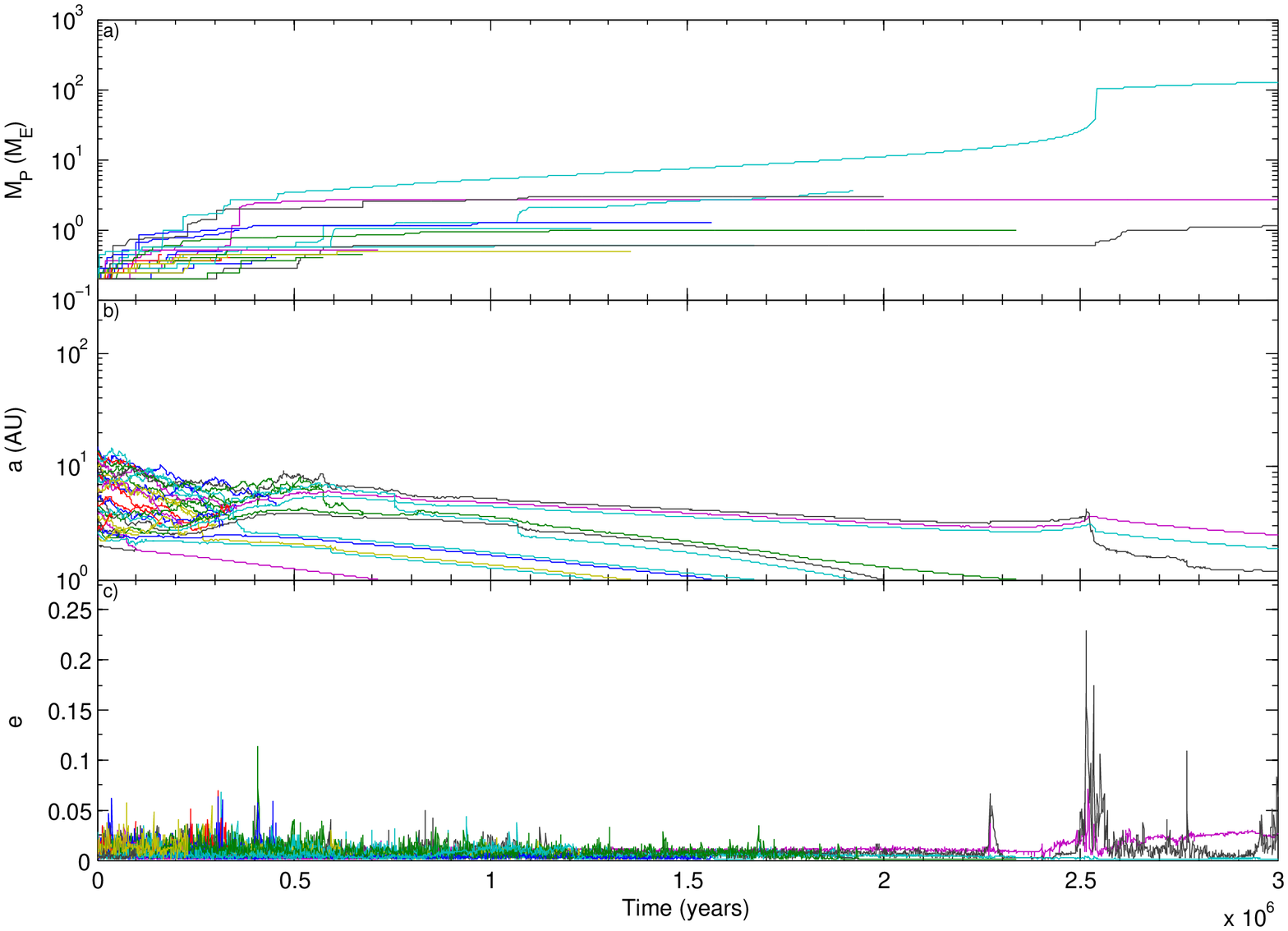}
      \caption{Evolution of masses, semimajor axes and eccentricities of all protoplanets
in simulation M03B-NA.\label{plot:M03B-NA_summary} }
    \end{center}
  \end{figure*}

\begin{figure*}
  \begin{center}
  \includegraphics[width=140mm, clip]{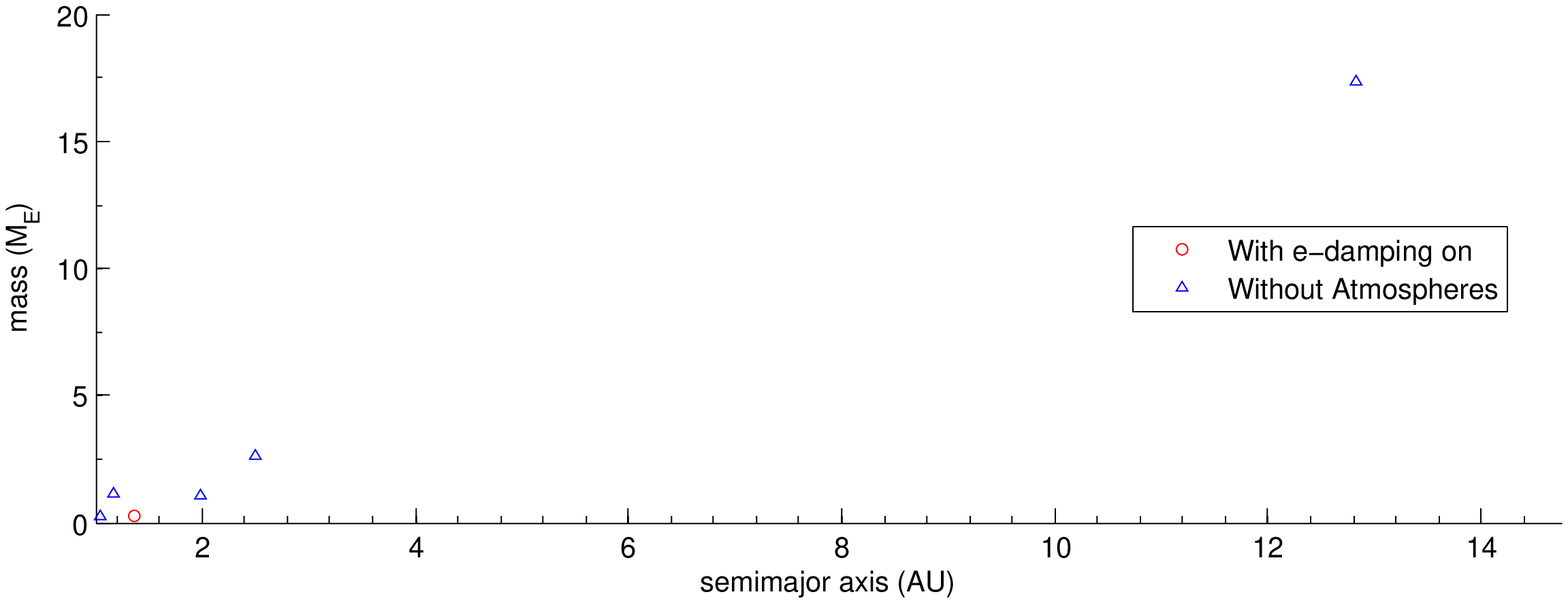}
  \caption{Summary of masses against semimajor axis for all small surviving planets outside 
1 AU for both runs where eccentricity damping was turned off and where enhanced capture radii due to 
atmospheric drag were turned off.\label{plot:naandedampsurvivors} }
  \end{center}
  \end{figure*}
  
    \subsection{Eccentricity modulation of corotation torques}
  \label{ecutoff}
Recent numerical simulations \citep{BitschKley2010} indicate that corotation torques are
substantially reduced in their effectiveness when a planet develops an eccentric
orbit. In particular, we expect the corotation torque to be effectively quenched when
the radial excursion associated with the eccentric orbit exceeds the width of the 
horseshoe region. In order to simulate this effect, we have run a few simulations 
(labelled 'E' in Table \ref{simparam}) where the eccentricity modulation function
in Eq.~\ref{eq:E} is switched on.
The effect of this was as one might expect: growth was significantly stunted compared 
to the corresponding runs without this reduction factor (runs M05A/B compare to eccentricity 
damping runs E01A/B and M06A/B compare to E02A/B). Nearly all protoplanets were lost beyond 
the inner edge by approximately 1 Myr and most protoplanets were lost within half this time.
Only one planet survived to run completion out of all four of the simulations and its final 
position is shown in Fig.~\ref{plot:naandedampsurvivors}, which shows a summary of surviving 
planets from the `E' runs, as well as those discussed below in which the enhanced planetesimal 
capture radii are switched off. A plot showing the time evolution of this particular simulation 
is given in Fig.~\ref{plot:E02B_summary_e_xs}, where we have plotted the eccentricity in
the bottom panel as $e/x_s$. It is clear that planet-planet interactions maintain 
values of $e/x_s \ge 1$ throughout the simulation, until it is depleted of planets through
their inward migration. This result suggests that closer investigation of the role of
planetary eccentricity in regulating the strength of corotation torques needs to be undertaken,
since the modest evidence we have accumulated suggests that mutual encounters between planets
may remove the benefits provided by corotation torques in enhancing the formation and
survival of planets. Similar conclusions have been reached in a recent study
by McNeil \& Nelson (In preparation) that examines the formation of hot Neptunes
and super-Earth planets in radiatively inefficient discs.
   
  \subsection{Capture radii enhancement switched off}
  \label{atmospheresoff}
A common outcome within our simulations has been the rapid formation and growth of planetary cores,
and their subsequent rapid migration inward. One reason for this rapid growth is our
adoption of an enhanced capture radius for planetesimals arising because of a
putative gaseous envelope settling onto protoplanets during their formation.
We re-ran the simulations described in Sect.~\ref{indruns} without enhanced capture radii.
Growth was notably slower as expected in all four runs. Two runs, however,
(corresponding to the M03B and M07B non-atmosphere runs)
did manage to form one gas giant planet each.

The time evolution of run M03B-NA (non-atmosphere) is shown in Fig.~\ref{plot:M03B-NA_summary}.
A planet grows slowly to just over 3 M$_{\oplus}$ by approximately 500,000 years. It sits in an 
area of the disc largely cleared of solid material by other protoplanets and slowly accretes gas 
before eventually undergoing runaway gas accretion at 2.5 Myrs and ends up at 2 AU with a mass of 
126 M$_{\oplus}$.

Run M07B-NA forms a gas giant by means of three 1-1.5 M$_{\oplus}$ bodies migrating out to large 
semimajor axes, and then merging to form a 3.5 M$_{\oplus}$ body at 40 AU which slowly accretes gas 
until runaway gas accretion occurs at 2.8 Myrs. The planet ends up at 50 AU with a mass of 
319 M$_{\oplus}$.
The lower mass planets that survived in these runs are shown in Fig.~\ref{plot:naandedampsurvivors}. 
Interestingly, one of these is a Neptune-sized planet.
  
  \section{Comparison with observations}
  \label{realobs}
  
    \begin{figure}
    \caption{Mass vs semimajor axis plot comparing observed extrasolar planets with 
our simulation results.
      \label{plot:summarytoreal}}
    \includegraphics[width=80mm, clip]{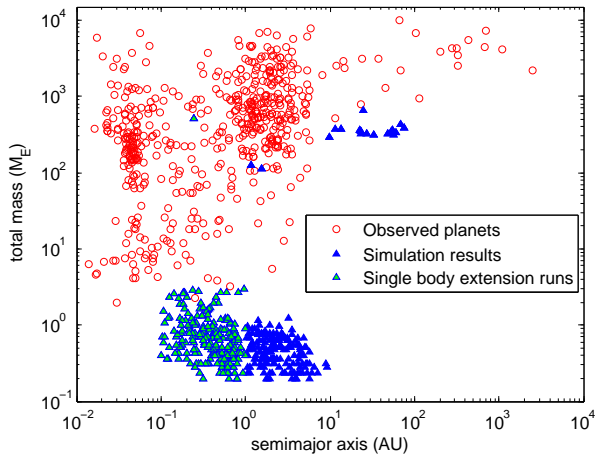}
  \end{figure}
 
The work presented in this paper is not intended to be a serious attempt at
planetary population synthesis modelling, unlike the work presented by
\cite{IdaLin2008} and \cite{Mordasini2009b}. Instead it is aimed at 
exploring the consequences of having strong corotation torques operating
during the oligarchic growth stage of planetary systems formation,
and understanding how planetary growth, migration and planet-planet interactions
combine to form planetary systems.
Nonetheless, it is of interest to examine how well the simple models that
we have presented compare with the observational data on extrasolar planets.
 
Figure \ref{plot:summarytoreal} shows a mass-period diagram with
our results overlaid on all current observed exoplanets (sourced from www.exoplanet.eu).
Our shorter period giant planets lie well in the range of already detected exoplanets both in 
terms of mass and semimajor axis. Our longer period planets, however, lie in an area 
that is sparsely populated with observational results. It is worth noting, however, that
this region of parameter space is much more problematic for the detection of planets, as
observations rely on direct imaging methods rather than radial velocity or transit observations.

A clear failing of our results is in the formation of super-Earths and Neptune-mass bodies.
One reason for this appears to be the gas accretion routine that we have adopted,
that allows planets to accrete gas once their masses exceed 3 M$_{\oplus}$.
An additional issue is the adoption of atmosphere-induced enhanced capture radii
for planetesimal accretion onto protoplanets. The very rapid growth of planets due to this
often causes them to migrate rapidly toward the central star, an outcome that is
reduced in models where enhanced capture radii are not included.
These are issues that we will address in a future publication.

 \section{Discussion and Conclusions}
  \label{conclusions}
Although the models presented in this paper include a broad range of physical
processes relevant to planet formation (migration; planetary growth through 
mutual protoplanet collisions; accretion of planetesimals and gas;
planet-planet gravitational interactions), we have adopted a number of
assumptions and simplifications that inevitably affect the realism of
the simulations and their results. These include:
  
\begin{enumerate}
\item \emph{Simulation domain}. Even though we have modelled a relatively wide semimajor axis domain 
with our initial solid matter disc compared to previous N-body work on planetary formation, 
it is clear that accurate modelling of discs in which significant corotation torques arise
requires as wide a domain as possible. Protoplanets move significantly in the disc with some 
forming at 2-3 AU and migrating out to 70-80 AU. Similarly, planets migrate inward and ought 
to traverse the terrestrial planet region which we have not modelled. Planets forming in the 
terrestrial region might also migrate out into the regions that we have investigated. In short, 
the migration behaviour observed in the simulations presented in this paper indicates that all
regions of the disc are coupled during planet formation, and it is no longer sensible to
think in terms of a ``terrestrial planet region'' or a ``giant planet region''.
As such, a suitable model would involve a domain ranging from as far in as 0.1 AU out to 
approximately 50 AU. Such a simulation is beyond current modelling techniques because
of the required numbers of protoplanets and planetesimals, even for the method
presented by \cite{McNeilNelson2009, McNeilNelson2010} which utilises multiple time steps in a
parallel symplectic integrator. Instead, such global models of planetary formation are
probably going to require efficient use of modern GPU technology.

\item \emph{Gas disc model}. We currently model the gas disc as having fixed power-laws in surface density
and temperature, with the disc mass undergoing exponential decay to mimic its viscous and photoevaporative
evolution. In reality, the disc is heated by the central star and through local viscous dissipation, and
it cools through radiative emission. A significant improvement to the model that we are in the
process of implementing will be to evolve the disc surface density and temperature explicity using
a 1+1D numerical solution, as described in \cite{PapaloizouTerquem1999}, for example.
This approach will be similar to that described in \cite{FoggNelson2009} 
and references therein,
and will allow gap formation and type II migration to be simulated directly, along with 
UV photoevaporation of the disc during its final stages of dispersal.

\item \emph{Planetary atmosphere model and enhanced capture radii.}
As described in the preceding sections, rapid planetary growth is assisted by 
the enhanced accretion of planetesimals through implementation of a model
for planetary atmospheres that increases the effective accretion cross section
\citep{inaba03}. Although this model works well when accretion is dominated
by planetesimals, it is probably inaccurate when accretion includes 
giant impacts between protoplanets. In particular, a planetary atmosphere
can be completely liberated from a planet when it is impacted by a body whose
mass is similar to that of the atmosphere, and our implementation of the
atmosphere model does not account for this effect. The atmosphere model
would clearly be improved in its accuracy if it responded to 
giant impacts as well as planetesimal accretion rates.

\item \emph{Gas envelope accretion}. Our model for gas accretion during the formation of 
gas giants is very rudimentary, although it serves the purpose of allowing gas giant 
planets to form in our simulations. While a full accretion model for each planet similar to those
presented in \cite{pollack96} and \cite{movshovitz} would be ideal, this 
is computationally beyond the reaches 
of an N-body code that can model planetary systems formation over Myr time scales. 
However, there are 
improvements that can be made that will allow local conditions in the disc to influence the
gas accretion rate onto a planet. Coupled with a more sophisticated disc model that allows
explicit modelling of gap formation and gas accretion, such an approach would alleviate
the requiement to set an artifical upper limit for the planet mass that can form through
runaway gas accretion.

\end{enumerate}

We have presented the results of simulations that include recent torque 
formulae for type I migration (including Lindblad and corotation torques), 
with gas envelope accretion, enhanced capture radii due to gas atmospheres,
and planet-planet gravitational dynamics included. We have surveyed a range of disc 
models which all allow
for outward migration driven by corotation torques. The main results that we have obtained may be 
summarised as follows:
\begin{itemize}
\item Convergent migration of protoplanets, and the rapid accretion of planetesimals,
can cause the rapid growth of planetary cores to masses in excess of 10 M$_{\oplus}$
within 0.5 Myr in most disc models. This leads to rapid inward migration of these bodies,
driven by Lindblad torques, when the horseshoe libration time scale becomes significantly 
shorter than the thermal/viscous diffusion time scale and the corotation torques saturate.

\item Formation of planetary cores with a few Earth masses $\ge 0.5$ Myr after the
simulations are initiated can lead to their migration into the outer regions of the
disc (30 - 50 AU). Steady mass growth through gas accretion onto the planet
can counterbalance the slow inward migration that occurs as the gas disc mass reduces,
and long-term survival in the outer disc can lead to gas giant planet formation
there when runaway gas accretion ensues. This mode of giant planet formation was
found to be a common outcome in our simulations, especially those with disc
models that generate strong outward migration, leading to numerous gas giant
planets being formed between semimajor axes 10 - 60 AU. Models such as
these could potentially explain the long-period giant planets discovered
recently through direct imaging \citep{Marois, Kalas, Lagrange}.

\item Out of 40 simulations that used disc models whose masses were either 3 or 5 times more
massive than the MMSN, 19 gas giant planets were formed. Most of these are similar in mass
to Jupiter (in part because of the gas accretion prescription that was adopted in the models),
and are formed at large distances from the star. Short period Jovian mass planets were also
formed, however, along with a pair of Saturn-mass bodies at 
intermediate ($\sim 1$ AU) orbit distances.
These latter systems were formed in discs that generate weaker corotation torques than
those that tend to generate the longer-period giant planets.

\item Multiplanet systems containing more than one giant planet were found to be
a rare outcome (3 out of 40 simulations produced two giant planets each), and this
has the additional effect of producing systems with very small eccentricities
and inclinations due to the low rate of occurence of planet-planet scattering events. In fact,
the only planets to be formed with significant eccentricities were a pair of
closely separated Jovian-mass objects that underwent significant dynamical interaction.

\item Our simulations completely fail to produce super-Earth or Neptune-mass planets.
This appears to arise because of very rapid inward migration of planets that grow early
in the disc lifetime and undergo rapid inward migration, combined with the switching-on
of gas accretion that converts planets of intermediate mass into gas giants at later times.
Modification of the planetary atmosphere and gas accretion prescriptions will probably
result in more surviving planets with intermediate masses.
Numerous planets in the Earth mass range were formed in the simulations, however.
The `desert' of super-Earths and Neptunes is similar to that reported 
in the planetary population synthesis models of \cite{IdaLin2008}, and
occurs for much the same reasons as theirs (rapid gas accretion and migration).

\item Simulations performed where the corotation torque is attenuated
when planet eccentricities grow to become larger than the dimensionless
horseshoe width appear to produce results quite different from those
in which this effect is neglected (i.e. all the runs described above). In particular
the growth and survival of planets is reduced because mutual encounters
between protoplanets maintains the typical eccentricities above the critical
value for which corotation torques diminish. In these latter simulations,
no gas giant planets were formed at all. Further work is required to establish the
influence of corotation torques on planet formation {\it via} oligarchic growth,
where planet-planet interactions maintain finite values of the eccentricity.

\end{itemize}

Our models demonstrate that strong corotation torques can substantially
alter the qualitative outcomes of planet formation simulations.
Even the simplest model of planet formation that involves non linear
gravitational interaction between protoplanets and planetesimals during
planetary accumulation is by its nature a chaotic process. Given a well-defined
distribution of initial protoplanet/planetesimal masses and orbital elements
from which individual formation models are drawn, however, an ensemble of such models 
should give rise to a distribution of outcomes with well-defined statistical 
properties. Allowing the set of initial conditions to be drawn from a range
of disc models whose life-times and radial profiles of density and temperature 
also have well-defined distributions serves only to modify the distribution of outcomes,
as does including additional physical processes such as type I migration.
A corollary of this argument is that increasing the complexity of migration
processes, as we have done in this paper, also serves to modify the distribution
of outcomes in a quantifiable manner. Corotation torques, however, increase
the dependency of formation outcomes on the details of the underlying disc model
and microphysical processes such as those that control the opacity of disc material.
The dependency of migration on opacity, turbulent viscosity and other disc properties,
and the role of migration in shaping planetary system architectures, increases
the need for more refined observations of protoplanetary disc properties
and improved disc models to allow planetary formation calculations to be
compared with data on extrasolar planetary systems in a meaningful way.
To summarise: planetary formation is a chaotic process, but is deterministic in a
statistical sense. Corotation torques do not change the validity of this statement,
however, their dependence on detailed disc properties increases the difficulty of 
constructing realistic planetary formation models for comparison with observational data.

This is the first in a series of papers to examine the oligarchic growth of planets
under the influence of type I migration, including corotation torques. Models that
include a more sophisticated treatment of the issues raised in Sect.~\ref{conclusions}
will be presented in a forthcoming publication.
    
  \section*{Acknowledgments}
P. Hellary is supported by a QMUL PhD studentship. The simulations presented in
this paper were performed on the High Performance Computing facility at QMUL
purchased under the SRIF and CIF initiatives. We acknowledge useful discussions
with D. McNeil concerning various aspects of this paper.

\label{lastpage}
  
\end{document}